\documentclass[cm]{aa}
\usepackage{rotating}
\usepackage{epsfig}

\begin{document}
\thesaurus{11.01.2; 11.09.3; 11.10.1; 13.18.1} 
\title{A new sample of giant radio galaxies from the WENSS survey}
\subtitle{II - A multi-frequency radio study of a complete sample: Properties of the radio lobes and their environment}
\author{A.P. Schoenmakers\inst{1,2,5}\thanks{{\it Present address: }NFRA, P.O. Box 2, 7990~AA Dwingeloo, The Netherlands}
\and K.-H. Mack\inst{3,4}
\and A.G. de Bruyn\inst{5,6}
\and H.J.A. R\"{o}ttgering\inst{2}
\and U. Klein\inst{4}
\and H. van der Laan\inst{1}} 
\institute{Astronomical Institute, Utrecht University, P.O. Box 80\,000,
3508~TA Utrecht, The Netherlands 
\and {Sterrewacht Leiden, Leiden University, P.O. Box 9513, 2300~RA Leiden, The Netherlands}
\and {Istituto di Radioastronomia del CNR, Via P. Gobetti 101, I-40129 Bologna, Italy}
\and {Radioastronomisches Institut der Universit\"at Bonn, Auf dem H\"ugel 71, D-53121 Bonn, Germany}
\and {NFRA, P.O. Box 2, 7990~AA Dwingeloo, The Netherlands}
\and {Kapteyn Astronomical Institute, University of Groningen, P.O. Box 800, 9700~AV Groningen, The Netherlands}}
\offprints{A.P. Schoenmakers (Schoenmakers@astron.nl)}

\date{Received ; accepted}
\titlerunning{Multi-frequency radio observations of GRGs}
\authorrunning{A.P. Schoenmakers et al.}
\maketitle

\begin{abstract}

We have formed a complete sample of 26 low redshift ($z \la 0.3$)
giant radio galaxies (GRGs) from the WENSS survey, selected at flux
densities above 1 Jy at 325 MHz. We present 10.5-GHz observations with
the 100-m Effelsberg telescope of 18 sources in this sample. These
observations, together with similar data of the remaining eight
sources, are combined with data from the WENSS, NVSS and GB6 surveys
to study the radio properties of the lobes of these sources at
arcminute resolution. We investigate radio source asymmetries,
equipartition energy densities in the lobes, the presence of lobe
pressure evolution with redshift, the spectral age and the density of
the environments of these sources. We find that the armlength
asymmetries of GRGs are slightly larger than those of smaller sized
3CR radio galaxies and that these are difficult to explain as arising
from orientation effects only.  We also find indications that the
lobes of the GRGs, despite their large sizes, are still overpressured
with respect to their environment. Further, we argue that any
evolution of lobe pressure with redshift in these large sources (e.g.
Cotter 1998) is due to selection effects. For sources which could be
used in a spectral ageing analysis, we find spectral ages which are
large, typically a few times $10^7$ yr. This is comparable to earlier
studies of some giant sources and indicates that such large spectral
ages are common for this class of radio source. The advance velocities
of the radio lobes are typically a few percent of the speed of light,
which is higher than those found for smaller, low power ($<10^{26.5}$
W Hz$^{-1}$ at 178 MHz) radio sources, and more comparable to higher
power radio sources. This suggests that the GRGs in our sample are the
oldest members of the group of relatively high power radio sources
whose radio powers have evolved to their currently observed lower
values (cf. Kaiser et al. 1997).
\end{abstract}

\begin{keywords}
Galaxies: active -- Intergalactic medium -- Galaxies: jets -- Radio continuum: galaxies
\end{keywords}

\section{Introduction}

Giant radio galaxies (GRGs) are the largest radio sources in the
Universe which are associated with active galactic nuclei (AGN). A
common definition for GRGs is that they are radio sources with a
linear size above 1 Mpc\footnote{We use $H_0 =
50$~km\,s$^{-1}$\,Mpc$^{-1}$ and $q_0 = 0.5$ throughout this
paper.}. These enormous sizes make them interesting objects to
study. Why are they so large? Is it because they grow much faster than
other radio galaxies, or are they extremely old radio sources? Which
are the properties of their progenitors? Also, because their size is
so extreme, they allow us to study their radio structures in detail
and to use them as probes of the gaseous environment of their host
galaxies on scales of a few hundred kpc to a few Mpc.

Since their discovery in the early seventies, several individual GRGs
have been the subject of detailed radio studies at a variety of
wavelengths and resolutions (e.g. 3C\,236 by Strom \& Willis 1980 and
Barthel et al. 1985; NGC\,315 by Willis et al. 1981, NGC\,6251 by
Perley et al. 1984). However, systematic studies of GRGs as a
population have always been hampered by the small number of sources
available and by non-uniform selection effects. Since GRGs are large
and their radio emission is not very powerful their surface brightness
is relatively low. This makes them difficult objects to detect or
recognize in most large-scale radio surveys. As a result, a large
fraction of the known GRGs have been discovered serendipitously
(e.g. Hine 1979, de Bruyn 1989), hence the difficulty in obtaining a
uniformly selected sample.  The most uniform dataset on GRGs available
yet is that of 10.5-GHz observations with the 100-m Effelsberg
telescope (Klein et al. 1994, Saripalli et al. 1996, Mack et
al. 1997). All sources have been observed in the same way and at
similar sensitivities ({\sc rms}-noise $\sim 1$ mJy beam$^{-1}$), so
that the results can be easily compared to each other.

In paper I (Schoenmakers et al. 2000a; see also Schoenmakers 1999) 
we present a new sample
of 47 GRGs selected from the 325-MHz WENSS survey (Rengelink et
al. 1997).  In this paper, we will define a complete subsample of 26
sources with a 325-MHz flux density above 1 Jy. This is the largest
complete sample of GRGs, with well understood selection effects (see
paper I), yet compiled. We have used this sample for several follow-up
studies, among which a study of their radio properties using
multi-frequency radio data.  For this purpose, we have obtained new
10.5-GHz radio data of 18 of these sources using observations with the
100-m Effelsberg telescope; the 8 remaining sources have already been
observed with this instrument (Klein et al. 1994, Saripalli et
al. 1996, Mack et al. 1997).  We have used these data to investigate
the high-frequency radio morphology, the magnetic field configuration
and, combined with data obtained at lower frequencies, the spectral
index distribution and spectral ages (e.g. Mack et al. 1998). The
analysis and results of this study are presented here.  Subsequent
papers will deal with the optical properties of the AGN and their
relation with the radio structure, and with the evolution of GRGs,
both in terms of cosmological evolution as in terms of intrinsic radio
source evolution.

In Sect. \ref{sec:the_sample} we present the complete sample of GRGs
and discuss some of its characteristics.  Section \ref{sec:radio_data}
presents the new 10.5-GHz radio data and lower frequency data for the
sources in the sample. In Sect. \ref{sec:asymmetry} we derive several
source asymmetry parameters and investigate the presence of
correlations between these. The ages and lobe advance velocities of
several GRGs are derived in Sect. \ref{sec:lobe_velocities}, and the
energy densities and lobe pressures are derived in
Sect. \ref{sec:energy_densities}. In Sect. \ref{sec:discussion} we
discuss the results, focusing on the spectral ages and the environment
of the GRGs. Our conclusions are presented in
Sect. \ref{sec:conclusions}.

Throughout this paper, a spectral index $\alpha$ is defined according
to the relation $S_{\nu} \propto \nu^{\alpha}$ between flux density
$S$ and frequency $\nu$. 

\section{A complete sample of GRGs}
\label{sec:the_sample}

The GRGs presented in paper I have been selected from the WENSS survey
using the criteria that they should have an angular size above
$5\arcmin$ and a distance from the galactic plane $\ge 12\fdg5$.  We
find in paper I that a WENSS radio source is most likely included in
the sample if $S_{int}/\theta_{max} \ga 0.025$~Jy/arcmin, where
$S_{int}$ is the integrated 325-MHz flux density, and $\theta_{max}$
the largest angular size of the radio source. We have called this the
sensitivity limit of our selection.
  
The total number of sources in the sample of paper I is 47, but at low
flux density levels (i.e. $\la 200$ mJy) several candidate sources
have not yet been identified, and some sources may have been missed
because they have not been recognized as single structures. Therefore,
we have selected a subsample of 26 sources on the basis of a flux
density at 325 MHz, $S_{325}\!>\!1$~Jy. At such high flux densities, it is
unlikely that a source has escaped detection or recognition as a GRG.

On basis of our sensitivity limit (see above), it is unlikely that a
1-Jy GRG will be recognized at a redshift below 0.014, due to its very
low surface brightness in that case. However, the lowest redshift
source in our sample is NGC\,315 at $z=0.0167$, so no source needed to
be excluded on basis of this limit.

We have omitted two giant FRI-type (Fanaroff \& Riley 1974) radio
sources from the sample: 3C\,31 (e.g. Strom et al. 1983) and HB\,13 
(e.g. Masson 1979). The considerations for doing so were the
following. First, the observed size of sources of this class depends
strongly on the surface brightness sensitivity of the radio
observations; only for edge-brightened FRII-type sources, the angular
size and source structure is reasonably well defined because of the
presence of hotspots and the usually better outlined radio lobe
morphology. The edge-darkened nature of FRI's makes it unlikely that
the 1-Jy sample is complete for FRI-type sources.  Second, the
properties of the radio lobes of FRI-type sources are known to be
different.  This is shown, for instance, by the different spectral
index distribution in the radio lobes (e.g. J\"{a}gers 1986, Parma et
al. 1999), which may indicate different mechanisms for the
acceleration and ageing of the radiating particles. 

There are a few remaining sources which are strictly of type FRI, but
also show properties commonly found in FRII-type sources, such as
hotspots or sharply bound lobe structures. Well known examples of such
sources are DA\,240, NGC\,315 and NGC\,6251. Since the measured size
of these sources is better constrained, we have left these sources in
the sample, although they are excluded from many of the analyses in
this paper.  They are indicated as type `FRI/II' in
Tab. \ref{tab:1-Jy-sample}.

Our 1-Jy sample therefore contains all non FRI-type sources with a
325-MHz flux density above 1 Jy, an angular size above $5\arcmin$, a
linear size above 1 Mpc, a declination above $+28\degr$, a distance
from the galactic plane $> 12.5\degr$ and a redshift above 0.014, in
total 26 sources.  Of these, 16 are GRGs which were previously known
and 10 are newly discovered GRGs (see paper I). The list of sources
and some of their properties are presented in
Tab. \ref{tab:1-Jy-sample}. The source B\,1918+516 has an 
uncertain redshift determination due to the faintness of its optical
host galaxy and the difficulty we had in actually identifying it; an
independent confirmation is needed.

\begin{table}[t!]
\setlength{\tabcolsep}{5.5pt}
\caption{\label{tab:1-Jy-sample} List of the 26 sources which form the
`1-Jy sample' of GRGs. Column 1 gives the name of the source in
IAU-notation, with coordinates in B\,1950.0; column 2 gives an
alternative, more common name, if available; column 3 indicates
whether it is one of the newly discovered GRGs (`N') or one of the
formerly known, or `old' GRGs (`O'); column 4 gives the redshift of
the source; column 5 gives the radio morphological type. A `B'
indicates it is a broad-line radio galaxy, i.e. the Hydrogen Balmer
lines have broad components; a `Q' indicates a quasar-like spectrum
(broad lines, blue continuum).  Column 6 gives the projected linear
size in Mpc. For references concerning the properties of the old GRGs
we refer to paper I.}
\begin{tabular}{l l l l l c}
\hline \hline \\
\multicolumn{1}{c}{(1)} & \multicolumn{1}{c}{(2)} &
\multicolumn{1}{c}{(3)} & \multicolumn{1}{c}{(4)} &
\multicolumn{1}{c}{(5)} & \multicolumn{1}{c}{(6)}  \\ 
IAU name & Alt. name & & \multicolumn{1}{c}{$z$} & \multicolumn{1}{c}{Type} & \multicolumn{1}{c}{$D$}  \\
 & & & & FR & \multicolumn{1}{c}{$[\,$Mpc$\,]$} \\
\hline \\             
B\,0050$+$402 &           & O & 0.1488    & II   & 1.5  \\
B\,0055$+$300 & NGC\,315  & O & 0.0167    & I/II & 1.7  \\
B\,0109$+$492 & 3C\,35    & O & 0.0670    & II   & 1.1  \\
B\,0136$+$396 & 4C\,39.04 & O & 0.2107    & II   & 1.6  \\
B\,0157$+$405 & 4C\,40.09 & O & 0.0827    & I/II & 1.9  \\
B\,0211$+$326 &           & N & 0.2605    & II   & 1.6  \\
B\,0309$+$411 &           & O & 0.1340    & II-B & 1.8  \\
B\,0648$+$733 &           & N & 0.1145    & II   & 1.9  \\
B\,0658$+$490 &           & N & 0.0650    & I/II & 1.9  \\
B\,0745$+$560 & DA\,240   & O & 0.0356    & I/II & 2.0  \\
B\,0813$+$758 &           & N & 0.2324    & II-B & 2.3  \\
B\,0945$+$734 & 4C\,73.08 & O & 0.0581    & II   & 1.5  \\
B\,1003$+$351 & 3C\,236   & O & 0.0989    & II   & 5.7  \\
B\,1209$+$745 & 4CT\,74.17& O & 0.107     & II   & 1.2  \\
B\,1213$+$422 &           & N & 0.2426    & II-B & 1.6  \\
B\,1309$+$412 &           & O & 0.1103    & II   & 1.0  \\
B\,1312$+$698 & DA\,340   & O & 0.106     & II   & 1.3  \\
B\,1358$+$305 &           & O & 0.206     & II   & 2.6  \\
B\,1426$+$295 &           & N & 0.0870    & II   & 1.9  \\
B\,1450$+$333 &           & N & 0.249     & II   & 1.7  \\
B\,1543$+$845 &           & N & 0.201     & II   & 2.1  \\
B\,1626$+$518 &           & O & 0.0547    & II   & 1.6  \\
B\,1637$+$826 & NGC\,6251 & O & 0.023     & I/II & 3.0  \\
B\,1918$+$516 &           & N & 0.284$^a$ & II   & 2.3  \\
B\,2043$+$749 & 4C\,74.26 & O & 0.104     & II-Q & 1.6  \\
B\,2147$+$816 &           & N & 0.1457    & II   & 3.7  \\
\hline \hline\\
\end{tabular}
\begin{minipage}{\linewidth}
Notes:\\
$a-$Redshift still uncertain (see text and paper I).
\end{minipage}
\end{table}

\section{Radio data and radio spectra of the 1-Jy sample}
\label{sec:radio_data}

In this section we present new 10.5-GHz radio data of 18 of the 26
sources in the 1-Jy sample of GRGs. We have measured total and lobe
flux densities of all sources at several frequencies between 325 MHz
and 10.5 GHz. Further, we present the radio spectra of the sources and
their radio lobes.

\subsection{The 10.5-GHz observations}

Eight sources in our sample were already observed at 10.5 GHz
with the 100-m Effelsberg telescope (Klein et al. 1994, Saripalli et
al. 1996, Mack et al. 1997). We have observed the 18 remaining sources 
with the same instrument and achieving a similar sensitivity of
$\sigma_{\rm rms}\!\approx\!1$ mJy beam$^{-1}$.

The observations have been carried out in multiple observing sessions
between December 1995 and April 1998, using the 4-horn receiver
operating at a central frequency of 10.45 GHz and employing a
bandwidth of 300 MHz.  The beam size at this frequency is $69\arcsec$
(FWHM).  For a detailed description of the basic observational
technique and data reduction procedure we refer to the paper by
Gregorini et al. (1992). The calibration of the flux density scale has
been achieved by mapping 3C\,286 and 3C\,295, with the flux density
scale adopted from Baars et al. (1977).  Field sizes and map centers
are compiled in Tab. \ref{tab:effelsberg-log}.  Mapping was performed
by scanning the telescope in azimuth with a speed of $40\arcmin$/min,
with a scan separation of $20\arcsec$ in elevation and using all four
horns. Difference maps have been computed from all horns to
efficiently suppress atmospheric disturbances of the signal.
Following the usual restoration technique of the differential maps
(Emerson et al. 1979) the maps were transformed into a right ascension
-- declination system.  All maps, with the exception of B\,0309$+$411,
have been CLEANed to remove sidelobes, applying the algorithm
described by Klein \& Mack (1995). During the observations of
B\,0309$+$411, snow was collected on the dish of the telescope. Since
this has affected the beam-shape, cleaning was not performed on this
source. Also, the snow led to problems with calibrating the
data. Therefore the flux density of B\,0309$+$411 is estimated to be
only correct to 20\%.  The individual coverages
(Tab. \ref{tab:effelsberg-log}) of each source have been averaged to
give the final Stokes' I, Q and U maps.  Contour plots of the radio
maps have been presented in Appendix A.

\begin{table*}[t]
\caption{\label{tab:effelsberg-log} Observing log of the 10.5-GHz observations.
Column 1 gives the name of the source. Columns 2 and 3 give the
B1950.0 coordinates of the center of the map. Column 4 gives the
number of coverages used in the final maps (i.e. after removing those
which suffered from bad atmospheric conditions). The number of
coverages used in the $I$, $Q$ and $U$ maps are the same. Column 5
gives the size of the area covered. Column 6 gives the observing dates
and the number of coverages on that date in the format `number of
coverages:month/year'. Column 7 and 8 give the {\sc rms}-noise in the
total power maps and in the $Q,U$ maps.}
\begin{tabular}{l l l c r@{$\times$}l l l c}
\hline \hline \\
\multicolumn{1}{c}{(1)} & \multicolumn{1}{c}{(2)} & \multicolumn{1}{c}{(3)} & \multicolumn{1}{c}{(4)} & \multicolumn{2}{c}{(5)} & \multicolumn{1}{c}{(6)} & \multicolumn{1}{c}{(7)} & \multicolumn{1}{c}{(8)} \\
\multicolumn{1}{c}{Source} & \multicolumn{1}{c}{R.A.} & \multicolumn{1}{c}{Decl.} & \multicolumn{1}{c}{Cov.}  & \multicolumn{2}{c}{Mapsize} & \multicolumn{1}{c}{Obs. dates} & \multicolumn{1}{c}{$\sigma_I$} & \multicolumn{1}{c}{$\sigma_{Q,U}$} \\
 & \multicolumn{1}{c}{$[$\,h~m~s\,$]$}& \multicolumn{1}{c}{$[\,\degr~\arcmin~\arcsec\,]$} & \multicolumn{1}{c}{} & \multicolumn{2}{c}{$[\,\arcmin\times\arcmin\,]$} &  \multicolumn{1}{c}{} &  \multicolumn{2}{c}{$[$\,mJy beam$^{-1}$\,$]$} \\
\hline \\
B\,0109$+$492 & 01 09 07.0 &  49 13 00.0  &  10  &  38 &18 & 11:12/95                                     & 1.1 & 0.3 \\
B\,0157$+$405 & 01 57 27.8 &  40 34 23.0  &  11  &  41 &20 & 1:05/94; 3:08/94; 2:11/94; 1:12/94; 4:12/95  & 1.3 & 0.3 \\
B\,0211$+$326 & 02 11 19.5 &  32 37 18.0  &  14  &  31 &10 & 14:12/96                 & 0.9 & 0.3 \\
B\,0309$+$411 & 03 09 49.9 &  41 08 33.0  &  13  &  34 &13 & 7:12/96; 8:01/97            & 1.1 & 0.3 \\
B\,0648$+$734 & 06 48 14.0 &  73 24 20.5  &  10  &  39 &18 & 10:12/95                 & 1.0 & 0.3 \\
B\,0658$+$490 & 06 58 34.4 &  49 01 33.0  &  11  &  45 &24 & 4:12/95; 12:12/96; 1:01/97           & 1.1 & 0.3 \\
B\,0813$+$758 & 08 13 35.3 &  75 48 53.0  &  \phantom{1}9  &  35 &14 & 10:12/96                 & 1.0 & 0.3 \\
B\,1209$+$745 & 12 09 32.0 &  74 36 34.1  &  \phantom{1}9  &  34 &13 & 9:12/96                  & 1.0 & 0.3 \\
B\,1213$+$422 & 12 13 38.8 &  42 16 17.0  &  \phantom{1}7  &  31 &10 & 7:12/96                  & 1.3 & 0.4 \\
B\,1312$+$698 & 13 12 25.7 &  69 52 55.0  &  \phantom{1}9  &  32 &11 & 9:12/96                  & 0.9 & 0.3 \\
B\,1426$+$295 & 14 26 09.0 &  29 30 53.0  &  12  &  43 &22 & 13:12/96                 & 1.1 & 0.3 \\
B\,1450$+$333 & 14 50 58.4 &  33 20 52.0  &  10  &  32 &11 & 10:04/98                 & 0.6 & 0.3 \\
B\,1543$+$845 & 15 43 54.9 &  84 32 25.0  &  11  &  33 &12 & 14:04/98                 & 0.6 & 0.3 \\
B\,1626$+$518 & 16 27 00.0 &  51 53 51.0  &  12  &  46 &25 & 12:12/95                 & 1.3 & 0.3 \\
B\,1918$+$516 & 19 18 09.6 &  51 37 30.0  &  \phantom{1}7  &  33 &12 & 4:12/96; 3:01/97            & 1.3 & 0.4 \\
B\,2147$+$816 & 21 46 48.2 &  81 40 11.0  &  11  &  45 &24 & 5:12/96; 5:04/98; 7:08/98         & 0.9 & 0.3 \\
\hline \hline\\
\end{tabular}
\end{table*}

\subsection{Other radio data}

\subsubsection{325-MHz data}

\begin{figure}[t!]
\resizebox{\hsize}{!}{\epsfig{file=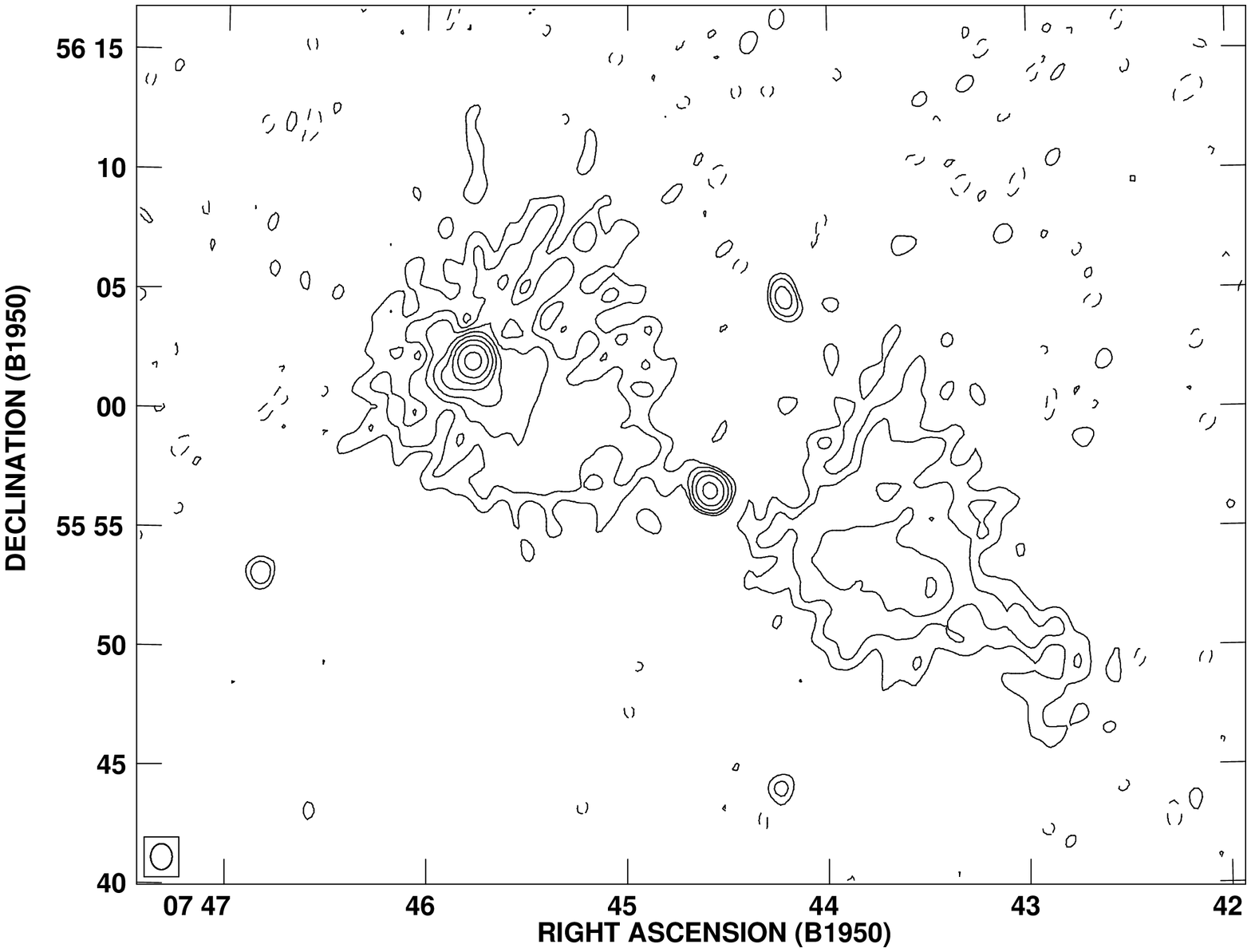,angle=270}} \\
\resizebox{\hsize}{!}{\epsfig{file=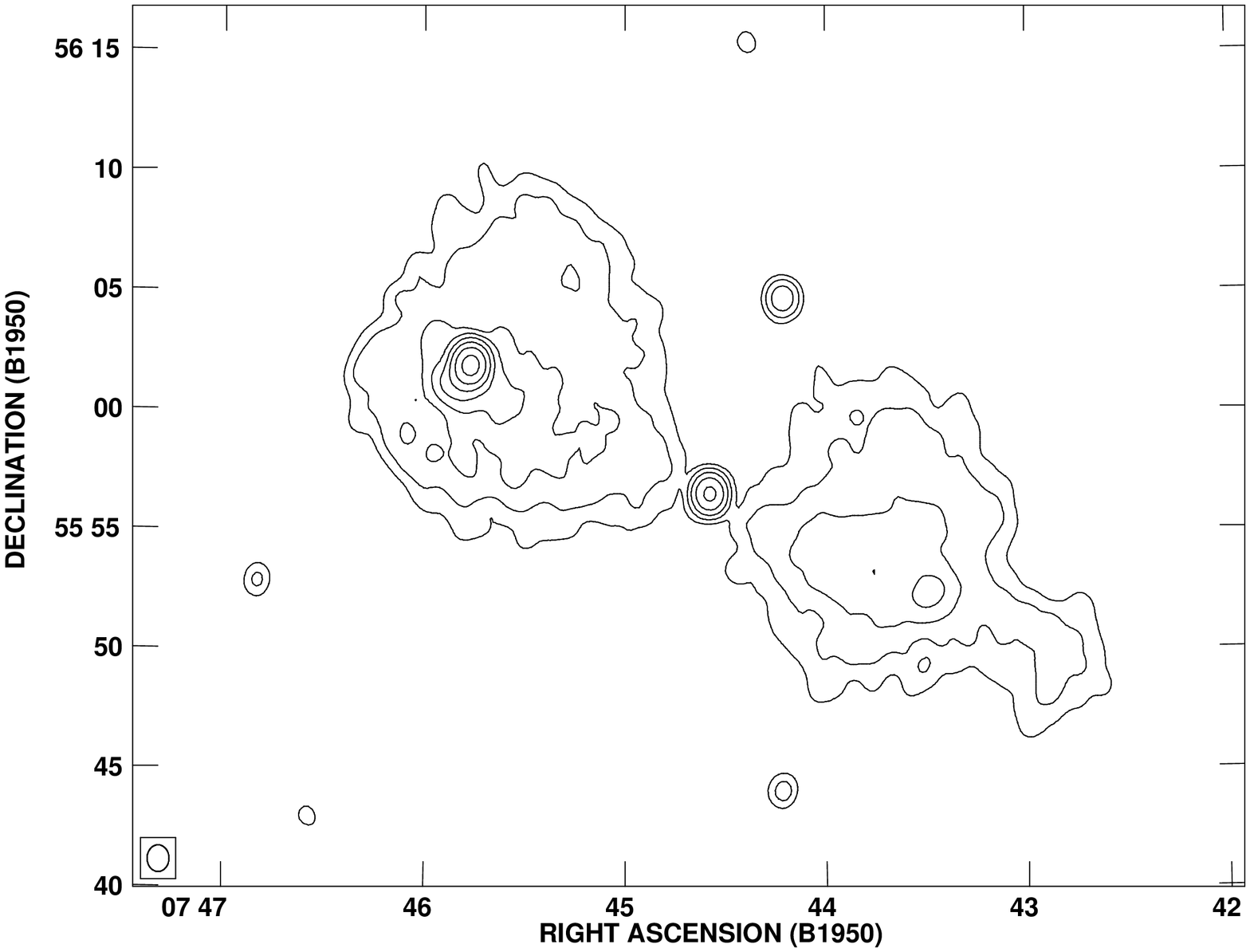,angle=270}}
\caption{\label{fig:DA240-comparison} Two contour plots of the giant
radio source DA\,240 at a frequency of 325 MHz. The upper plot is from
the WENSS survey. Contours are drawn at
$15\times(-1,1,2,4,8,16,32,64,128)$ mJy beam$^{-1}$. The lower plot is
from a 325-MHz WSRT observation by Mack et al. (1997). Contours are at
the same levels as in the top plot.}
\end{figure}

We have measured the 325-MHz flux densities and morphological
parameters on the radio maps of the WENSS survey.  The mosaicing technique
used in the observations for WENSS (see Rengelink et al. 1997 for a
description) results in a highly uniform coverage of the
$(u,v)$-plane, with baselines as short as $\sim 40\lambda$.  WENSS is
therefore potentially well suited to obtain accurate flux densities of
extended structures. However, since the $(u,v)$-plane is not as well
sampled as with continuous observations, in case of complicated and
very extended source structures WENSS cannot map all source components
reliably.  This is particularly notable in sources such as DA\,240 or
NGC\,6251.  To illustrate this problem we present in
Fig. \ref{fig:DA240-comparison} a map of the source DA\,240 as it
appears in the WENSS and as it is published by Mack et al. (1997) from
a complete 12-hr WSRT synthesis observation. In the latter case, not
only the noise level is much lower, but also the fainter extended
radio structures are much better reproduced. Since Mack et al. have
published maps resulting from full synthesis WSRT observations for the
four spatially largest objects in our sample (NGC\,315, DA\,240,
3C\,236, NGC\,6251), which wil suffer most from this effect, 
we have measured the flux densities using their maps.

\subsubsection{1.4-GHz data}

All sources have also been observed at 1.4 GHz in the NVSS survey (Condon
et al. 1998). The observations for the NVSS survey were done in a
`snap-shot' mode, using the VLA in its D-configuration with baselines
down to $\sim 170\lambda$ only. Both these aspects seriously degrade
the sensitivity for structures above 10\arcmin~-- 15\arcmin~in angular
size.  We therefore only present 1.4-GHz flux densities from the NVSS
survey for sources smaller than $10\arcmin$.

\subsubsection{4.85-GHz data}

Five GRGs in our sample (NGC\,315, DA\,240, 3C\,236, NGC\,6251 and
B\,1358+295) have been observed at 4.75 GHz with the 100-m Effelsberg
telescope (Parma et al. 1996; Mack et al. 1997). The beam size for
these observations is $\sim\!2\farcm5$ (FWHM). For sources without
5-GHz Effelsberg data we have used the 4.85-GHz Greenbank survey (GB6;
Gregory et al. 1996) to measure flux densities.  This survey has
mapped the northern sky up to a declination of $+75\degr$, using a
beam size of $3\farcm2 \times 3\farcm7$ (FWHM).  For sources above
$+75\degr$ declination, we have no 5-GHz flux densities. The flux
densities in the source catalogue of the GB6 survey were obtained by
fitting Gaussians to the observed sources, which, since all GRGs in
our sample are larger than the beam size of the GB6 survey, gives
unreliable results in our case.  Therefore, we have retrieved the
original digital {\sc fits}-format images, using the {\it SkyView}
database, and measured the flux densities directly from the maps.
Background offsets in the maps have been determined by measuring the
mean flux density in an area directly surrounding the source,
carefully omitting any significant sources in this area.  Discrete
sources which overlap with the radio structure of the GRGs have been
identified in the higher resolution NVSS and 10.5-GHz Effelsberg radio
maps. Their contribution to the measured 4.85-GHz flux density has been
subtracted by estimating their 4.85-GHz flux density using a power-law
interpolation of their 1.4 and 10.5-GHz flux densities.

\subsection{Radio spectra}
\label{sec:radio-spectra}

We have measured the total integrated flux densities, $S_{int}$, at
325 MHz, 4.8 GHz and 10.5 GHz. We have also measured the flux
densities of the lobes separately at 325 MHz, 1.4 MHz (for sources
below $10\arcmin$ in angular size), 4.8 GHz (for sources above
$10\arcmin$ in angular size and declination below $+75\degr$) and 10.5
GHz.  Only in the case of B\,0309$+$411 we have not measured the lobe
flux densities at 10.5 GHz due to the strongly dominating radio core
at that frequency. The flux densities have been tabulated in
Tab. \ref{tab:lobe-fluxes}.

\begin{figure*}
\centering
\resizebox{0.9\hsize}{!}{\epsfig{file=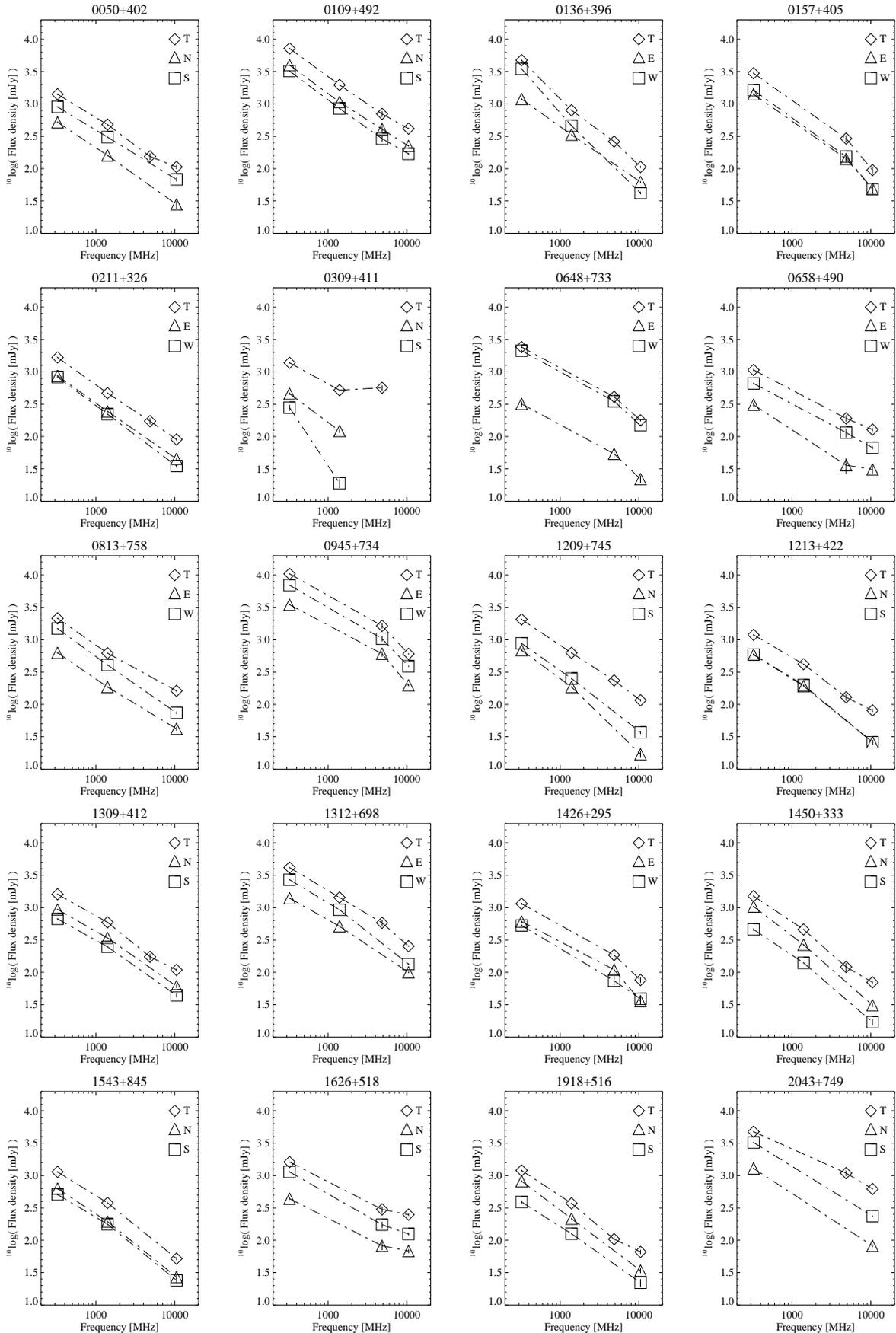}}
\caption{\label{fig:spectra_all} Radio spectra of the
GRGs and their radio lobes. For each source we have plotted the total
integrated flux density (indicated by `T'), and the integrated flux
density of the two lobes (indicated by `E',`W', etc.) where
available. The name of the source is indicated above each panel. See
the text for more details.}
\end{figure*}

The radio spectra of the sources with more than two flux density
measurements are plotted in Fig. \ref{fig:spectra_all}, based on the
flux densities from Tab. \ref{tab:lobe-fluxes}. We have used separate
signs for the total integrated flux densities and those of the two
lobes. Not all sources have been plotted here; similar radio spectra
of the sources B\,0055$+$300 (NGC\,315), B\,1003$+$351 (3C\,236),
B\,0745$+$560 (DA\,240) and B\,1637$+$826 (NGC\,6251) can be found in
Mack et al. (1997); for the source B\,1358$+$305 we refer to the paper
by Parma et al. (1996). The source B\,2147$+$815 has not been plotted
since we only have data at two frequencies for this source (see
Tab. \ref{tab:lobe-fluxes}).

For several sources the spectrum of the total integrated emission clearly
steepens towards higher frequencies (e.g. B\,0157$+$405,
B\,0648$+$733, B\,0945$+$734 and B\,1312$+$698). This is usually a
sign of spectral ageing of the radiating particles in the source.  In
other cases the spectrum appears to flatten (e.g. B\,0309$+$411,
B\,1626$+$518). Since all these sources have bright radio cores at
10.5 GHz, this must be the result of the radio core having a flat, or
inverted, spectrum.

\begin{sidewaystable*}
\centering
\setlength{\tabcolsep}{4.5pt}
\caption{\label{tab:lobe-fluxes} Flux densities of the GRGs and their
components. Column 1 gives the name of the source. Column 2 indicates
which side of the source is named $A$ and $B$ in this table (`N'
stands for north, etc.). Columns 3 to 5 give the integrated flux
density, $S$, at 325 MHz of the whole source and of each of the two
sides of the source. Columns 6 through 8 give the flux densities at
1.4 GHz for sources with an angular size below $10\arcmin$. Column 9
through 11 give the flux densities at 4.85 GHz for sources with
declination below $+75\degr$. For sources larger than $10\arcmin$,
also the flux densities of the lobes have been measured. Columns 12
through 14 give the flux densities at 10.5 GHz.}
\begin{small}
\begin{tabular}{l c r@{$\,\pm\,$}r r@{$\,\pm\,$}r r@{$\,\pm\,$}r r@{$\,\pm\,$}r r@{$\,\pm\,$}r
r@{$\,\pm\,$}r r@{$\,\pm\,$}r r@{$\,\pm\,$}r r@{$\,\pm\,$}r r@{$\,\pm\,$}r r@{$\,\pm\,$}r r@{$\,\pm\,$}r}
\hline \hline \\
\multicolumn{1}{c}{(1)} & \multicolumn{1}{c}{(2)} & \multicolumn{2}{c}{(3)} & \multicolumn{2}{c}{(4)} & \multicolumn{2}{c}{(5)} & \multicolumn{2}{c}{(6)} & \multicolumn{2}{c}{(7)} & \multicolumn{2}{c}{(8)} &
 \multicolumn{2}{c}{(9)} & \multicolumn{2}{c}{(10)} & \multicolumn{2}{c}{(11)}  & \multicolumn{2}{c}{(12)} & \multicolumn{2}{c}{(13)} & \multicolumn{2}{c}{(14)}\\
 & & \multicolumn{6}{c}{\hrulefill~325 MHz~\hrulefill} & \multicolumn{6}{c}{\hrulefill~1.4 GHz~\hrulefill} & \multicolumn{6}{c}{\hrulefill~4.8 GHz~\hrulefill} & \multicolumn{6}{c}{\hrulefill~10.5 GHz~\hrulefill} \\
Source & \multicolumn{1}{c}{A,B} & \multicolumn{2}{c}{$S_{int}$} & \multicolumn{2}{c}{$S_A$}  & \multicolumn{2}{c}{$S_B$} & \multicolumn{2}{c}{$S_{int}$}  & \multicolumn{2}{c}{$S_A$} & \multicolumn{2}{c}{$S_B$} & \multicolumn{2}{c}{$S_{int}$} & \multicolumn{2}{c}{$S_A$} & \multicolumn{2}{c}{$S_B$} & \multicolumn{2}{c}{$S_{int}$} & \multicolumn{2}{c}{$S_A$} & \multicolumn{2}{c}{$S_B$} \\
 & & \multicolumn{2}{c}{$[$\,Jy\,$]$} & \multicolumn{2}{c}{$[$\,Jy\,$]$} & \multicolumn{2}{c}{$[$\,Jy\,$]$} &  \multicolumn{2}{c}{$[$\,mJy\,$]$} &  \multicolumn{2}{c}{$[$\,mJy\,$]$} &  \multicolumn{2}{c}{$[$\,mJy\,$]$} &  \multicolumn{2}{c}{$[$\,mJy\,$]$} & \multicolumn{2}{c}{$[$\,mJy\,$]$} &  \multicolumn{2}{c}{$[$\,mJy\,$]$} & \multicolumn{2}{c}{$[$\,mJy\,$]$} &  \multicolumn{2}{c}{$[$\,mJy\,$]$} & \multicolumn{2}{c}{$[$\,mJy\,$]$}\\[1ex]   
\hline \\
B\,0050$+$402     & N,S  &  1.41 & 0.04 & 0.52 & 0.02 & 0.90 & 0.02 &  480 & 10 &  160 &  4 &  307 &  7 &  154 & 19 & \multicolumn{4}{c}{\ }                &  106 &  8 &  28 &  4 &  68 &  5\\
B\,0055$+$300$^a$ & E,W  &  9.71 & 0.18 & 2.53 & 0.06 & 3.55 & 0.10 &   \multicolumn{2}{c}{\ }   &  \multicolumn{4}{c}{\ }               & 2460 & 70 &  296 & 40 &  569 & 50 & 2056 & 50 & 271 & 50 & 646 & 80\\
B\,0109$+$492     & N,S  &  7.19 & 0.15 & 3.98 & 0.08 & 3.22 & 0.06 & 1965 & 40 & 1067 & 21 &  851 & 19 &  702 & 67 &  407 & 40 &  290 & 30 &  415 & 11 & 224 &  6 & 169 &  5\\
B\,0136$+$396     & E,W  &  4.75 & 0.10 & 1.19 & 0.03 & 3.48 & 0.07 &  798 & 16 &  334 &  7 &  460 & 10 &  262 & 30 &  \multicolumn{4}{c}{\ }               &  106 &  4 &  62 &  2 &  42 &  2\\ 
B\,0157$+$405     & E,W  &  2.98 & 0.08 & 1.41 & 0.04 & 1.63 & 0.04 &   \multicolumn{2}{c}{\ }   &  \multicolumn{4}{c}{\ }               &  292 & 35 &  141 & 17 &  153 & 18 &   95 & 11 &  49 &  8 &  48 &  8\\ 
B\,0211$+$326     & E,W  &  1.67 & 0.04 & 0.87 & 0.02 & 0.83 & 0.02 &  469 & 10 &  245 &  5 &  222 &  5 &  173 & 18 &  \multicolumn{4}{c}{\ }               &   90 &  4 &  45 &  2 &  35 &  2\\  
B\,0309$+$411$^b$ & N,S  &  1.38 & 0.04 & 0.46 & 0.02 & 0.28 & 0.03 &  520 & 11 &  122 &  9 &   19 &  3 &  567 & 50 &  \multicolumn{2}{c}{\ } & 550 & 110 & \multicolumn{2}{c}{\ } \\
B\,0648$+$733     & E,W  &  2.41 & 0.06 & 0.32 & 0.03 & 2.12 & 0.05 &   \multicolumn{2}{c}{\ }   &  \multicolumn{4}{c}{\ }               &  411 & 40 &   54 &  9 &  351 & 33 &  178 &  7 &  22 &  3 & 149 &  4\\  
B\,0658$+$490$^c$ & E,W  &  1.07 & 0.04 & 0.31 & 0.02 & 0.66 & 0.02 &   \multicolumn{2}{c}{\ }   &  \multicolumn{4}{c}{\ }               &  191 & 22 &   36 & 10 &  115 & 19 &  128 &  8 &  31 &  5 &  67 &  4\\
B\,0745$+$560$^a$ & E,W  & 17.05 & 0.35 &10.21 & 0.21 & 6.07 & 0.13 &   \multicolumn{2}{c}{\ }   &  \multicolumn{4}{c}{\ }               & 1780 & 50 & 1187 & 23 &  457 & 20 & 1041 & 47 & 625 & 20 & 165 & 15\\
B\,0813$+$758     & E,W  &  2.14 & 0.05 & 0.63 & 0.02 & 1.49 & 0.03 &  621 & 13 &  186 &  5 &  408 &  9 &  \multicolumn{2}{c}{\ }    &  \multicolumn{4}{c}{\ }               &  162 &  6 &  42 &  3 &  74 &  3\\ 
B\,0945$+$734     & E,W  & 10.43 & 0.21 & 3.48 & 0.08 & 6.96 & 0.14 &   \multicolumn{2}{c}{\ }   &  \multicolumn{4}{c}{\ }               & 1641 &150 &  607 & 57 & 1037 & 95 &  605 & 15 & 198 &  7 & 389 & 10\\  
B\,1003$+$351$^a$ & E,W  & 13.13 & 0.26 & 2.18 & 0.08 & 2.99 & 0.06 &   \multicolumn{2}{c}{\ }   &  \multicolumn{4}{c}{\ }               & 2200 & 45 &  213 & 20 &  401 & 20 & 1208 & 28 & 177 & 10 & 166 & 10\\ 
B\,1209$+$745     & N,S  &  2.06 & 0.05 & 0.69 & 0.02 & 0.88 & 0.02 &  627 & 13 &  186 &  4 &  252 &  7 &  235 & 25 &  \multicolumn{4}{c}{\ }               &  116 &  5 &  17 &  2 &  37 &  2\\ 
B\,1213$+$422     & N,S  &  1.19 & 0.03 & 0.59 & 0.02 & 0.59 & 0.02 &  419 &  9 &  192 &  4 &  199 &  4 &  129 & 15 &  \multicolumn{4}{c}{\ }               &   81 &  5 &  26 &  2 &  26 &  2\\  
B\,1309$+$412     & N,S  &  1.61 & 0.04 & 0.94 & 0.02 & 0.67 & 0.02 &  594 & 12 &  341 &  7 &  248 &  5 &  174 & 20 &  \multicolumn{4}{c}{\ }               &  109 &  5 &  61 &  4 &  44 &  3\\ 
B\,1312$+$698     & E,W  &  4.16 & 0.09 & 1.40 & 0.03 & 2.70 & 0.06 & 1436 & 29 &  515 & 11 &  928 & 19 &  583 & 55 &  \multicolumn{4}{c}{\ }               &  254 &  7 & 100 &  3 & 134 &  4\\ 
B\,1358$+$305$^d$ & N,S  &  1.84 & 0.04 & 1.32 & 0.03 & 0.52 & 0.02 &  451 & 10 &  318 &  7 &  136 &  4 &  122 &  4 &   95 &  4 &   28 &  4 &   52 &  2 &  34 &  2 &  19 &  2\\
B\,1426$+$295$^e$ & E,W  &  1.15 & 0.04 & 0.61 & 0.02 & 0.53 & 0.02 &   \multicolumn{2}{c}{\ }   &  \multicolumn{4}{c}{\ }               &  185 & 36 &  110 & 31 &   74 & 11 &   76 &  7 &  36 &  4 &  39 &  4\\  
B\,1450$+$333     & N,S  &  1.51 & 0.04 & 1.04 & 0.03 & 0.46 & 0.02 &  460 & 10 &  267 &  6 &  140 &  3 &  121 & 18 &  \multicolumn{4}{c}{\ }               &   70 &  3 &  31 &  2 &  17 &  2\\  
B\,1543$+$845     & N,S  &  1.14 & 0.03 & 0.63 & 0.02 & 0.51 & 0.02 &  378 &  9 &  194 &  4 &  177 &  4 &  \multicolumn{2}{c}{\ }    &  \multicolumn{4}{c}{\ }               &   52 &  3 &  27 &  2 &  24 &  2\\
B\,1626$+$518$^f$ & N,S  &  1.62 & 0.03 & 0.44 & 0.02 & 1.14 & 0.03 &   \multicolumn{2}{c}{\ }   &  \multicolumn{4}{c}{\ }               &  300 & 31 &   82 & 12 &  174 & 19 &  249 & 11 &  68 &  5 & 125 &  6\\  
B\,1637$+$826$^g$ & E,W  & 11.55 & 0.23 & 2.70 & 0.09 & 4.01 & 0.10 &   \multicolumn{2}{c}{\ }   &  \multicolumn{4}{c}{\ }               &  \multicolumn{2}{c}{\ }    &  \multicolumn{4}{c}{\ }               & 1558 & 40 &  85 & 20 & 216 & 15\\ 
B\,1918$+$516     & N,S  &  1.20 & 0.03 & 0.82 & 0.02 & 0.39 & 0.02 &  372 &  8 &  215 &  5 &  126 &  3 &  104 & 14 &  \multicolumn{4}{c}{\ }               &   66 &  5 &  34 &  3 &  22 &  3\\  
B\,2043$+$749$^h$ & N,S  &  4.76 & 0.10 & 1.29 & 0.03 & 3.24 & 0.07 &   \multicolumn{2}{c}{\ }   &  \multicolumn{4}{c}{\ }               & 1091 &105 &  \multicolumn{4}{c}{\ }               &  621 & 14 &  82 &  4 & 236 &  6\\  
B\,2147$+$816     & N,S  &  1.06 & 0.05 & 0.46 & 0.02 & 0.56 & 0.02 &   \multicolumn{2}{c}{\ }   &  \multicolumn{4}{c}{\ }               &  \multicolumn{2}{c}{\ }    &  \multicolumn{4}{c}{\ }               &  111 &  7 &  47 &  4 &  52 &  4\\ 
\hline \hline \\
\end{tabular}
\begin{minipage}{\linewidth}
Notes: $a-$4.75-GHz flux densities from Mack et al. (1997); $b-$Lobe
flux densities measured after subtracting unresolved source at core
position. No 10.5-GHz lobe flux densities given because of strong core
dominance at this frequency. $c-$4.85-GHz flux density of eastern lobe
after subtracting the estimated core flux density (by interpolation of
the 1.4 and 10.5-GHz flux density); $d-$4.75-GHz flux densities from
Parma et al. (1996); $e-$At 4.85 GHz (GB6), confused with bright source
of estimated flux density of 165 mJy (by power-law interpolation of
the 1.4 and 10.5 GHz flux density); $f-$4.85-GHz flux density of
southern lobe after subtracting the estimated core flux density (by
power-law interpolation of the 1.4 and 10.5-GHz flux density);
$g-$Flux densities of the western lobe exclude the core-jet
structure. Also, no 4.8-GHz data available; $h-$No lobe flux densities
measured at 4.85 GHz due to the dominating radio core.\\
\end{minipage}
\end{small}
\end{sidewaystable*}

\section{Radio source asymmetries}
\label{sec:asymmetry}

In this section we measure the asymmetries in armlength, flux
density, spectral index, etc. We compare the values and
correlations we find for the GRGs with results obtained for samples of
smaller-sized sources.

\subsection{Morphological asymmetries}
\label{sec:morphology}

Asymmetries in the morphology of a radio source are common and may
reflect asymmetries in their environment. McCarthy et al. (1991), for
instance, find that in a sample of powerful 3CR radio sources there is
a correlation between the side of the source with the shortest arm and
the side of the source with the highest optical emission-line flux. If
a larger intensity of the line emission is due to a higher amount of
gas, then the advance of the radio lobe on that side of the source may
have been slowed down. On the other hand, the filling factor of this
gas may be too low to play an important role in the dynamical
evolution of the radio lobe (e.g. Begelmann \& Cioffi 1989). Also, at
large distances (i.e. $\ga 100$ kpc) from the host galaxy and the AGN,
this warm gas is difficult to detect by emission-line studies even if
it were abundant, so that identifying the side with the highest gas
density around the lobes becomes extremely difficult for larger
sources.  If line emitting gas or clouds are dynamically unimportant
for large sources, then the armlength asymmetries of radio galaxies
may reflect asymmetries in the distribution of the hot ($\sim 10^7$ K)
diffuse IGM around the host galaxy. In principle, GRGs thus allow us
to investigate the uniformity of the IGM on scales up to a few Mpc,
which is well outside of the reach of current X-ray instruments, apart
from a few very luminous clusters.

However, armlength asymmetries can also be a result of orientation
effects (e.g. Best et al. 1995; Scheuer 1995), since the forward edges
of the two lobes have different light travel times to the
observer. Best et al. (1995) explain the differences in the armlength
asymmetries between radio galaxies and quasars in a sample of 3CR
sources with orientation differences only, although high expansion
velocities of the lobes, up to $0.4c$, are then required. Since Best et
al. find no significant correlation between the parameters describing
the arm-length asymmetry and the emission-line asymmetry, they suggest
that environmental effects are not necessarily the main cause of the
observed asymmetries in 3CR sources, in agreement with Begelmann \&
Cioffi (1989).

To investigate if the linear size of the radio source has any
influence on the observed asymmetries, we have measured the armlengths
of the lobes of all FRII-type sources in the 1-Jy sample. 
We have calculated the armlength-ratio, $Q$, by dividing the length of
the longest arm by that of the shortest arm. This yields the
fractional separation difference, $x$, which is defined as $x =
\frac{Q-1}{Q+1}$ (e.g. Best et al. 1995). The advantage of using $x$,
instead of $Q$, is that its range is limited between 0 and 1.  In
Tab. \ref{tab:morphologies} we present the armlengths, $Q$, $x$, and
the references to the data used to measure these parameters.

\begin{figure}[bt]
\resizebox{\hsize}{!}{\epsfig{file=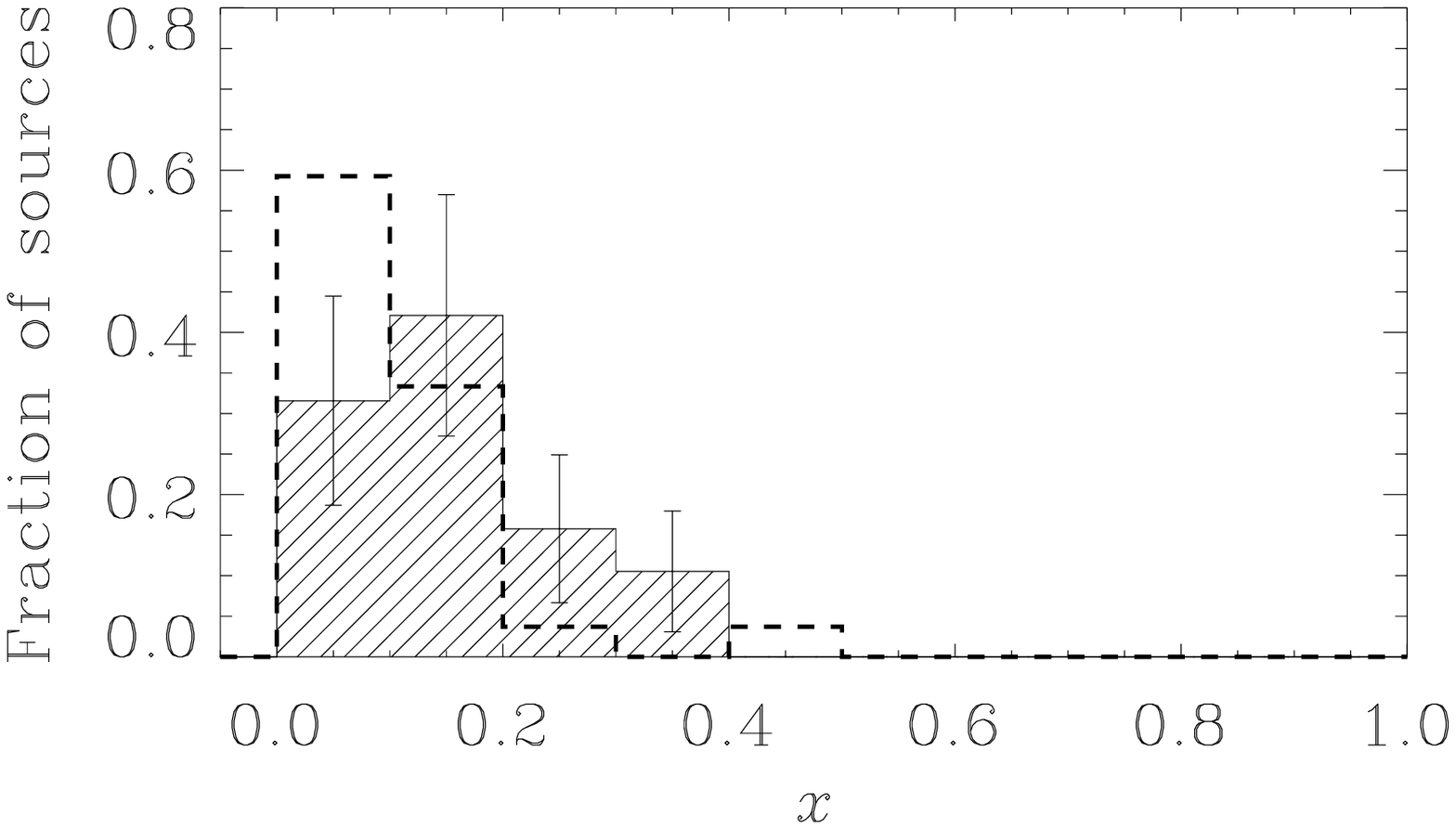}}\\
\resizebox{\hsize}{!}{\epsfig{file=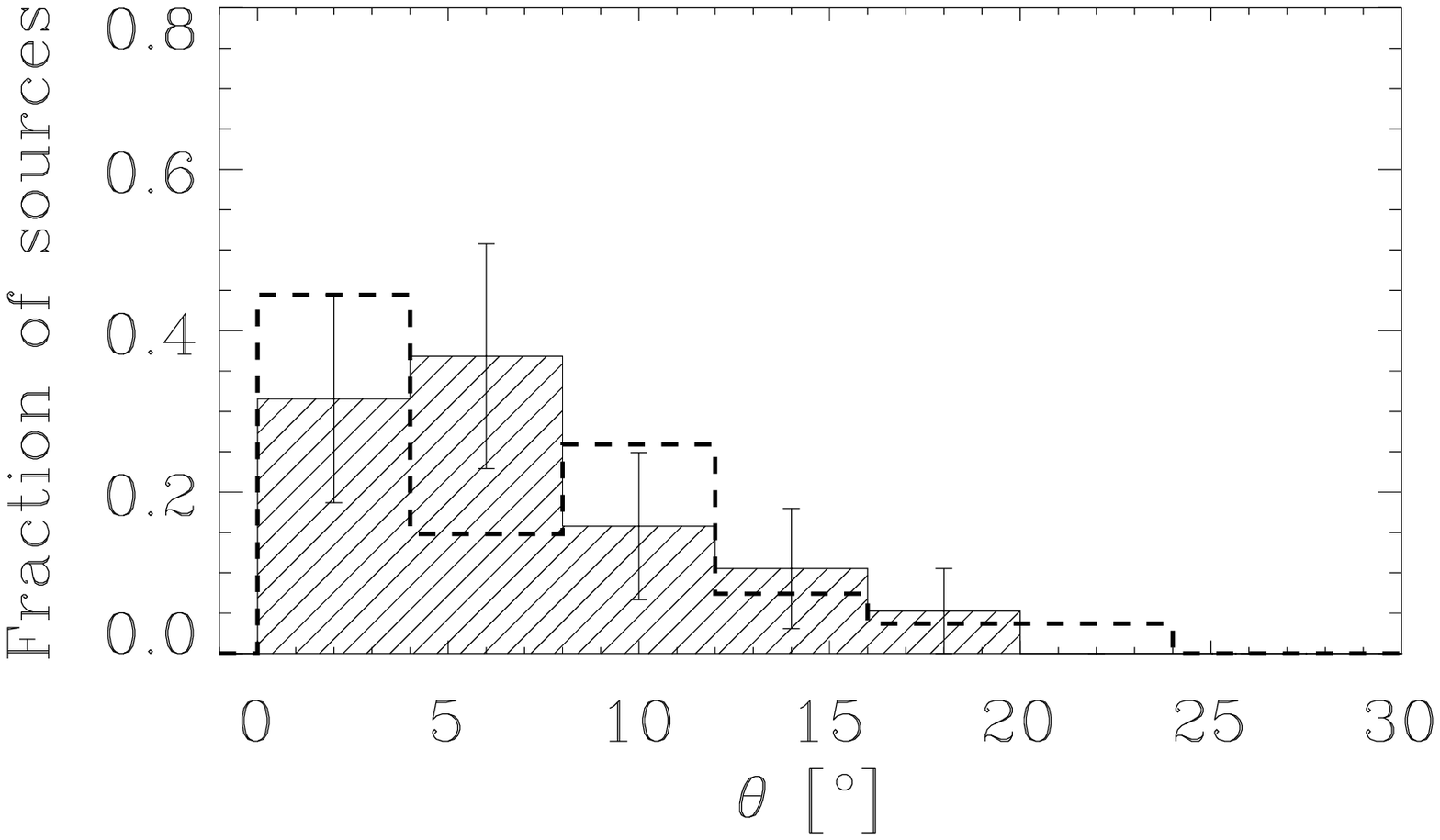}}
\caption{\label{fig:asymmetry} Plots of the distribution of the
fractional separation difference, $x$ (top), and the bending angle,
$\theta$ (bottom), of the FRII-type GRGs (hatched area) 
and of $z<0.3$ and $50 < D < 1000$~kpc powerful 3CR
sources from Best et al. (1995; area under dashed line). The error bars
assume a poisson distribution of the number of
sources in a bin.}
\end{figure}

In Fig. \ref{fig:asymmetry}a we have plotted a histogram of the
fractional separation difference of the GRGs. We have omitted the
source B\,2043$+$745 from our GRG sample since it is identified with a
quasar and may thus have an extreme orientation; note, however, that
this source is very symmetrical ($x=0.01 \pm 0.01$) and that including
it would raise the first bin only by $\sim 0.05$.

As comparison we have plotted the armlength asymmetry distribution of
$z<0.3$ (i.e. similar redshift range) FRII-type 3CR radio galaxies
with $50\!<\!D\!<\!1000$ kpc, for which we have taken the data from
Best et al. (1995). We have removed sources smaller than 50 kpc since
their asymmetries, if environmental, more reflect the gas distribution
inside or close to the host galaxy whereas we are interested in the
large-scale environment.  There are 27 sources in the 3CR subsample,
as compared to 19 FRII-type GRGs.

This comparison is only meaningful if 3CR sources are in similar gaseous
environments as GRGs, and will develop into GRGs provided that their
nuclear activity lasts for a long enough 
time. As yet, there is little detailed knowledge on the difference in
the environments of
3CR and GRG sources, and on the evolution of radio sources from small
to large ones (e.g. Schoenmakers 1999). 
Radio source evolution models (e.g. Kaiser et al. (1997) and
Blundell et al. (1999)) predict that 
GRGs must have been much more radio luminous when they were of smaller
size, which is not inconsistent with them being 3CR galaxies at an
earlier evolutionary stage (see also Schoenmakers 1999). We 
will therefore assume that the environments are largely similar
for the $>50$ kpc 3CR sources and the GRGs.

Although the difference in armlength asymmetry between 3CR radio
galaxies and GRGs is small and probably not significant, the GRGs
tend to be biased towards higher armlength asymmetries
(Fig. \ref{fig:asymmetry}a). A Kolmogorov-Smirnoff (K-S) test shows
that the two distributions are different at the 95\% confidence
level. Note, however, the relatively large (Poissoneous) errors in
Fig. \ref{fig:asymmetry}a, due to the small number of sources 
in the samples.

We have also measured the bending angle, defined as the angle between the
lines connecting the core with the endpoints of the two lobes. The
results are presented in Tab. \ref{tab:morphologies}. The distribution
of bending angles is plotted in Fig. \ref{fig:asymmetry}b, together
with the values for the $z<0.3$ 3CR galaxies from Best et
al. (1995). The distributions are quite similar; a K-S test shows that
they do not differ significantly at the 90\% confidence level.
Further discussion of these asymmetries will be presented in
Sect. \ref{sec:homogeniety}.

\begin{table*}[t]
\centering
\caption{\label{tab:morphologies} Morphological parameters of the
radio lobes. Column 1 gives the name of the source. Column 2 indicates
which side of the source is named $A$ and $B$ in this table (`N'
stands for north, etc.). Columns 3 to 6 give the angular size and the
physical size of the lobes. Column 7 gives the asymmetry parameter,
$Q$, defined as ratio of the length of the longest lobe to that of the
shortest. Column 8 gives the fractional separation parameter, $x$,
defined as $x = (Q-1)/(Q+1)$.  Column 9 gives the bending angle,
$\theta$, of the radio source, defined as the angle between the radio
axes of the two lobes.}
\begin{tabular}{l c r@{$\,\pm\,$}r r@{$\,\pm\,$}r r@{$\,\pm\,$}r r@{$\,\pm\,$}r r@{$\,\pm\,$}r r@{$\,\pm\,$}r r@{$\,\pm\,$}r}
\hline \hline\\
\multicolumn{1}{c}{(1)} & \multicolumn{1}{c}{(2)} & \multicolumn{2}{c}{(3)} & \multicolumn{2}{c}{(4)} & \multicolumn{2}{c}{(5)} & \multicolumn{2}{c}{(6)} & \multicolumn{2}{c}{(7)} & \multicolumn{2}{c}{(8)} &
 \multicolumn{2}{c}{(9)}\\
Source & \multicolumn{1}{c}{A,B} & \multicolumn{4}{c}{\hrulefill~$D_A$~\hrulefill} & \multicolumn{4}{c}{\hrulefill~$D_B$~\hrulefill}  & \multicolumn{2}{c}{$Q$} & \multicolumn{2}{c}{$x$} & \multicolumn{2}{c}{$\theta$} \\
 & & \multicolumn{2}{c}{$[$\,\arcsec\,$]$} & \multicolumn{2}{c}{$[$\,kpc\,$]$} & \multicolumn{2}{c}{$[$\,\arcsec\,$]$} &  \multicolumn{2}{c}{$[$\,kpc\,$]$} &  \multicolumn{4}{c}{\ } &  \multicolumn{2}{c}{$[$\,$\degr$\,$]$} \\
\hline \\
B\,0050$+$402 & N,S &  203 &   4 &  688 &  13 &  254 &   4 &  861 &  13 & 1.25 & 0.03 & 0.11 & 0.01 &  3 & 1 \\
B\,0109$+$492 & N,S &  323 &   2 &  561 &   3 &  323 &   2 &  561 &   3 & 1.00 & 0.01 & 0.00 & 0.00 &  7 & 1 \\
B\,0136$+$396 & E,W &  201 &   8 &  880 &  35 &  164 &   8 &  718 &  35 & 1.23 & 0.08 & 0.10 & 0.03 &  0 & 2 \\
B\,0211$+$326 & E,W &  167 &   5 &  841 &  25 &  140 &   5 &  705 &  25 & 1.19 & 0.06 & 0.09 & 0.02 &  3 & 2 \\
B\,0648$+$733 & E,W &  420 &   3 & 1155 &   8 &  320 &   3 &  880 &   8 & 1.31 & 0.02 & 0.14 & 0.01 & 10 & 1 \\
B\,0813$+$758 & E,W &  316 &  40 & 1478 & 187 &  185 &   3 &  865 &  14 & 1.71 & 0.22 & 0.26 & 0.06 & 13 & 2 \\
B\,0945$+$734 & E,W &  370 &   3 &  566 &   4 &  593 &  10 &  907 &  15 & 1.60 & 0.03 & 0.23 & 0.01 &  5 & 2 \\
B\,1003$+$351 & E,W & 1450 &  10 & 3533 &  24 &  980 &  10 & 2388 &  24 & 1.48 & 0.02 & 0.19 & 0.01 &  4 & 2 \\
B\,1209$+$745 & N,S &  280 &  20 &  728 &  52 &  162 &   5 &  421 &  13 & 1.73 & 0.13 & 0.27 & 0.04 & 18 & 3 \\
B\,1213$+$422 & N,S &  168 &   3 &  808 &  14 &  141 &  10 &  678 &  48 & 1.19 & 0.09 & 0.09 & 0.04 &  0 & 2 \\
B\,1309$+$412 & N,S &  186 &   3 &  496 &   8 &  186 &   3 &  496 &   8 & 1.00 & 0.02 & 0.00 & 0.01 &  0 & 2 \\
B\,1312$+$698 & E,W &  352 &   5 &  909 &  12 &  174 &   5 &  449 &  12 & 2.02 & 0.06 & 0.34 & 0.01 &  7 & 2 \\
B\,1358$+$305 & N,S &  200 &   8 &  862 &  34 &  448 &   8 & 1930 &  34 & 2.24 & 0.10 & 0.38 & 0.02 &  6 & 2 \\
B\,1426$+$295 & E,W &  365 &  10 &  797 &  21 &  530 &  10 & 1158 &  21 & 1.45 & 0.05 & 0.18 & 0.02 &  7 & 2 \\
B\,1450$+$333 & N,S &  154 &   5 &  754 &  24 &  193 &   5 &  945 &  24 & 1.25 & 0.05 & 0.11 & 0.02 & 12 & 2 \\
B\,1543$+$845 & N,S &  230 &   6 &  974 &  25 &  260 &   6 & 1101 &  25 & 1.13 & 0.04 & 0.06 & 0.02 &  0 & 2 \\
B\,1626$+$518 & N,S &  680 &  15 &  984 &  21 &  464 &  20 &  672 &  28 & 1.47 & 0.07 & 0.19 & 0.02 &  8 & 2 \\
B\,1918$+$516 & N,S &  180 &   5 &  957 &  26 &  260 &   5 & 1383 &  26 & 1.44 & 0.05 & 0.18 & 0.02 &  5 & 1 \\
B\,2043$+$749 & N,S &  450 &   5 & 1143 &  12 &  460 &   5 & 1169 &  12 & 1.02 & 0.02 & 0.01 & 0.01 &  9 & 2 \\
B\,2147$+$816 & N,S &  580 &  10 & 1935 &  33 &  520 &  10 & 1734 &  33 & 1.12 & 0.03 & 0.05 & 0.01 & 11 & 2 \\
\hline \hline\\
\end{tabular}
\end{table*}

\subsection{Flux density and spectral index asymmetries}
\label{sec:flux-asymmetries}

We have measured the flux density asymmetry, $R$, of the radio lobes
and the spectral index difference, $\Delta\alpha$, between the two
lobes.  We have defined these parameters such that they increase
monotonically with increasing asymmetry, i.e. $R$ is the 325-MHz flux
density of the brightest lobe divided by that of the weakest lobe and
$\Delta\alpha$ is the spectral index of the lobe with the flattest
spectrum minus the spectral index of the lobe with the steepest
spectrum, measured between 325 MHz and 10.5 GHz. We have searched for
correlations between these parameters and the armlength asymmetry
parameter $x$ using Spearman rank correlation tests. To avoid
spuriously significant correlations as a result of single outliers in
the parameter space under investigation, we have omitted, for each of
the two parameters being tested, the source with the highest value of
that parameter. The results of the correlation tests are presented in
columns (1) -- (3) of Tab. \ref{tab:rank-correlations}.
 
\begin{figure}[tb]
\resizebox{\hsize}{!}{\epsfig{file=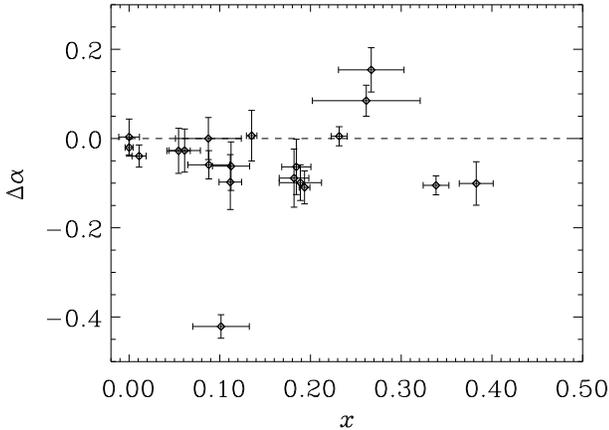,angle=90}}
\caption{\label{fig:sidif_x} The difference in spectral index between
the longest and the shortest lobe against the armlength asymmetry
parameter $x$ of the GRGs. The source with the largest side-to-side
difference in spectral index ($\Delta\alpha \approx -0.4$) is B\,0136+396.}
\end{figure}

We find that the only significant correlation is that between $x$ and
$\Delta\alpha$, i.e. when a radio source is more asymmetric in
armlength, then also the spectral index difference between the lobes
is systematically larger.  We find that in 15 out of 20 sources the
radio lobe with the shortest arm preferentially has a steeper
spectrum, although the difference in spectral index between the two
sides is $\la 0.1$ for almost all sources (see
Fig. \ref{fig:sidif_x}).

The lobe flux density asymmetry, $R$, is not significantly correlated
with either $x$ or $\Delta\alpha$. Still, we find that in 13 out of 20
sources (i.e. 65\%) the most luminous radio lobe has the shortest armlength. 
This is a similar percentage as found in the sample of 3CR sources
studied by McCarthy et al. (1991, but see Best et al. 1995) which
suggests that this trend does not occur by chance, only.

Since the luminosity of a radio lobe is the lobe volume integrated
emissivity, it is perhaps preferable to compare $R$ and $\Delta\alpha$
with the asymmetry in the estimated volume of the radio lobes. Also,
if the asymmetries are caused by large-scale environmental
inhomogeneities, this will probably affect the dynamical evolution of
the lobe as a whole, and not only its forward advance. Therefore, we
have investigated the correlation of $R$ and $\Delta\alpha$ with
$Q_V$, the ratio of the volume of the largest lobe to that of the
smallest. See Sect. \ref{sec:en-profiles} for the method used to
estimate the lobe volumes.

The results are presented in the last three columns of
Tab. \ref{tab:rank-correlations}.  
Although the correlation analysis gives significant results for both
$\Delta\alpha$ and $R$ with $Q_V$, indicating a correlated increase in
asymmetry for these parameters, these results are not very meaningful. 
We find that in only 12 out of 20 sources the
lobe which is smallest in estimated volume has the steepest
spectrum. Also, in only 11 out of 20 sources we find that the largest
lobe is the brightest. This indicates that the relative volume of a radio
lobe has less influence on its relative spectral index than the
armlength of the lobe. We will discuss this in more detail in
Sect. \ref{sec:homogeniety}.

\begin{table}
\caption{\label{tab:rank-correlations}
Spearman rank correlation tests between the asymmetry parameters
of the GRGs. See the text for details on the definition of the
parameters. Column 1 and 4 give the parameters being tested. Columns 2
and 5 give the correlation coefficients, $r_s$. Columns 3 and 6 give
the significance, $s$, of the correlation. The probability of the
correlation occurring by chance is $1-s$.}
\begin{tabular}{r r r r r r}
\hline \hline\\[-1ex]
\multicolumn{1}{c}{(1)} & \multicolumn{1}{c}{(2)} & \multicolumn{1}{c}{(3)} & \multicolumn{1}{c}{(4)} & \multicolumn{1}{c}{(5)} & \multicolumn{1}{c}{(6)}\\
 & \multicolumn{1}{c}{$r_s$} & \multicolumn{1}{c}{$s$} & & \multicolumn{1}{c}{$r_s$} & \multicolumn{1}{c}{$s$} \\ 
\hline\\[-1ex]
$\Delta\alpha - x$ & 0.657 & 0.997 & $\Delta\alpha - Q_V$ & 0.704 & 0.999 \\
$R - x$            & 0.257 & 0.664 & $R - Q_V$            & 0.582 & 0.986 \\
$\Delta\alpha - R$ & 0.333 & 0.809 \\
\hline \hline \\
\end{tabular} 
\end{table}

\section{The advance velocities and ages of the radio lobes}
\label{sec:lobe_velocities}

In this section we derive the ages and velocities of the lobes of the
GRGs by making use of the often observed trend that the radio spectrum
in the radio lobes of FRII-type sources steepens along the line
connecting the hotspots with the core.  This can be explained as a
result of ageing of the radiating particles after they have been
accelerated in the hotspots (e.g. Alexander \& Leahy 1987).  By fitting
spectral ageing models, the spectral index as a function of distance
from the hotspot then yields the age as a function of hotspot. From the
velocity of the hotspot and the size of the source an estimate of the
spectral age of the radio source is obtained.  Using ram-pressure
equilibrium at the head of the jet, we can then also estimate the
density in the ambient medium of the radio lobes.

\subsection{Spectral index profiles}
\label{sec:si-profiles}

We have used the WENSS or 325-MHz WSRT maps of Mack et al. (1997), and the 10.5-GHz Effelsberg data to
produce profiles of the spectral index, $\alpha_{325}^{10500}$, of the
GRGs along their radio axes. 
We have selected the 15 FRII-type sources with an angular size above
$7\arcmin$, only. This ensures that the radio lobes are covered by
several independent beams. We have omitted the source B\,1918$+$516
since it is confused with two unresolved sources in our low frequency
maps (see paper I).  We
have convolved the maps at 325/354 MHz to the resolution of the 10.5-GHz
Effelsberg maps ($69\arcsec \times 69\arcsec$ FWHM). In case that the
declination of a source is below $+51\degr$, we have convolved both
the 325/354-MHz and the 10.5-GHz maps to a common resolution of $69\arcsec
\times (54\cdot{\rm cosec}\,\delta)\arcsec$ (FWHM).  Additionally, for
sources with an angular size $\le 10\arcmin$ and for which reliable
1.4-GHz NVSS maps are therefore available, we have also made profiles
of the spectral indices between 325/354 MHz and 1.4 GHz (B\,0050$+$402,
B\,0813$+$758, B\,1312$+$698, B\,1543$+$845), or between 1.4 GHz and
10.5 GHz (B\,1358$+$305; B\,0050$+$402 was omitted here because only few
reliable datapoints could be found). 
For this, the NVSS maps have been convolved 
to the resolution of either the WENSS or the Effelsberg observations.

We have superposed rectangular boxes on the radio source, with the long
side oriented perpendicular to the radio axis and with a width along
the radio axes of $0.5$ times the
FWHM beam size of the convolved map.  The whole array of such boxes
has been centered on the radio core position. Possible confusing
sources have been blanked from the radio maps. 
In each box, we have integrated the flux density at each frequency and
calculated the spectral index of the box from this. If the flux density
inside a box was not significant (i.e. $<\! 3\sigma_I$) we used a
$3\sigma_I$ upper flux density limit to limit the spectral index.

The observations we use
have been done using different instruments (i.e. VLA, WSRT, 100-m
Effelsberg) which may lead to artefacts
in the spectral index profiles. In general, these artefacts are most prominent
in areas of low signal-to-noise (background level effects) and in areas with large intensity
gradients (beam-size effects), such as along the edge of a radio
source. Boxes with a low
signal-to-noise have been presented as limits in the
spectral index profiles, so background effects should not be serious in these
areas. Large intensity gradients occur mostly at the outer edges of the
sources, near the hotspots. Therefore, the outermost point on each side
of a source is usually not reliable.  

\begin{figure*}[t!]
\setlength{\tabcolsep}{0pt}
\begin{tabular}{l l l}
\resizebox{0.33\hsize}{!}{\epsfig{file=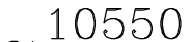,angle=90}} & 
\resizebox{0.33\hsize}{!}{\epsfig{file=0109+492.siplot.eps,angle=90}} &
\resizebox{0.33\hsize}{!}{\epsfig{file=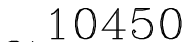,angle=90}} \\  
\resizebox{0.33\hsize}{!}{\epsfig{file=0813+758.si_highplot.eps,angle=90}} &
\resizebox{0.33\hsize}{!}{\epsfig{file=0945+734.siplot.eps,angle=90}} & 
\resizebox{0.33\hsize}{!}{\epsfig{file=1003+351.siplot.eps,angle=90}} \\
\resizebox{0.33\hsize}{!}{\epsfig{file=1209+745.si_highplot.eps,angle=90}} &
\resizebox{0.33\hsize}{!}{\epsfig{file=1312+698.si_highplot.eps,angle=90}} & 
\resizebox{0.33\hsize}{!}{\epsfig{file=1358+305.si_highplot.eps,angle=90}} \\
\resizebox{0.33\hsize}{!}{\epsfig{file=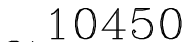,angle=90}} &
\resizebox{0.33\hsize}{!}{\epsfig{file=1543+845.siplot_high.eps,angle=90}} &
\resizebox{0.33\hsize}{!}{\epsfig{file=1626+518.siplot.eps,angle=90}} \\
\resizebox{0.33\hsize}{!}{\epsfig{file=2043+749.siplot.eps,angle=90}} & 
\resizebox{0.33\hsize}{!}{\epsfig{file=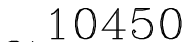,angle=90}}  \\
\end{tabular}
\caption{\label{fig:si-profiles} Profiles of the spectral index
distribution along the major axes of the radio sources, between 325
(or 354) MHz and 10.5 GHz.  The zero point on the x-axis is the
position of the radio core. The numbers along the upper axis denote
the distance from the radio core in units of kpc. Negative
values are on that side of the source which is mentioned first in
column two of Tab. \ref{tab:energy-densities}.}
\end{figure*}

\begin{figure*}[t!]
\setlength{\tabcolsep}{0pt}
\begin{tabular}{l l l}
\resizebox{0.33\hsize}{!}{\epsfig{file=0050+402.siplot.eps,angle=90}} & 
\resizebox{0.33\hsize}{!}{\epsfig{file=0813+758.siplot.eps,angle=90}} &
\resizebox{0.33\hsize}{!}{\epsfig{file=1209+745.siplot.eps,angle=90}} \\
\resizebox{0.33\hsize}{!}{\epsfig{file=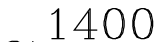,angle=90}} & 
\resizebox{0.33\hsize}{!}{\epsfig{file=1358+305.siplot.eps,angle=90}} & 
\resizebox{0.33\hsize}{!}{\epsfig{file=1543+845.siplot.eps,angle=90}} \\
\end{tabular}
\caption{\label{fig:si-profiles2} Profiles of the spectral index
distribution along the major axes of the sources smaller than
$10\arcmin$, between 325 (or 354) MHz and 1.4 GHz (except for
B\,1358+305, which is between 1.4 and 10.5 GHz).  The axes have been annotated similarly as in Fig. \ref{fig:si-profiles}.}
\end{figure*}

The profiles of the spectral index between
325/354 MHz and 10.5 GHz are presented in Fig. \ref{fig:si-profiles},
those between 325/354 MHz and 1.4 GHz in Fig. \ref{fig:si-profiles2}.
We find that only a few of the GRGs show a significant steepening of their radio
spectra towards the core (e.g. B\,2043$+$749). In other sources, such as
B\,2147$+$816, such a behaviour is not observed. In some cases a likely
explanation is that a flat-spectrum radio jet
contributes to the extended lobe emission. The source B\,1209$+$745 is
known to have a prominent one-sided jet pointing towards the north
(e.g. van Breugel \& Willis 1981), which explains the flat-spec\-trum
`plateau' north of the radio core position in
Fig. \ref{fig:si-profiles}. 
The same may be true for B\,1312$+$698, which shows jet-like features
in an (unpublished) 1.4-GHz WSRT radio map. However, for other sources such a scenario is
less probable since no jet-like features appear in any of our radio
maps. Among the possible causes for the apparent absence of ageing in these sources can be mixing of the lobe
material due to backflows in the lobes, non-uniform magnetic fields and
changes in the energy distribution of the accelerated particles during
the sources lifetime.

\subsection{The advance velocities of the radio lobes}
\label{sec:velocities}
 
\begin{figure*}[tb]
\setlength{\tabcolsep}{0pt}
\begin{tabular}{l l}
\resizebox{0.5\hsize}{!}{\epsfig{file=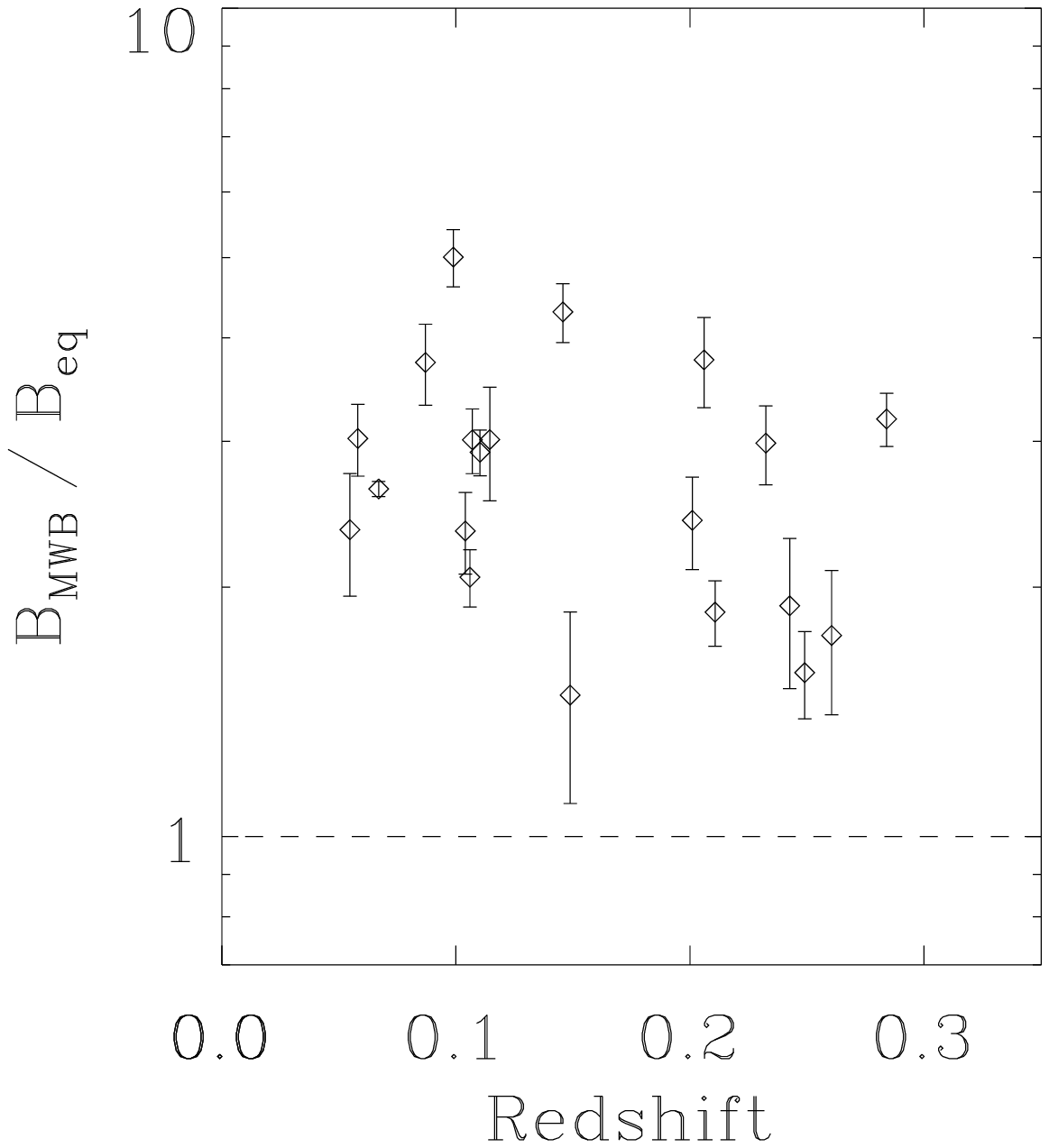}} & 
\resizebox{0.5\hsize}{!}{\epsfig{file=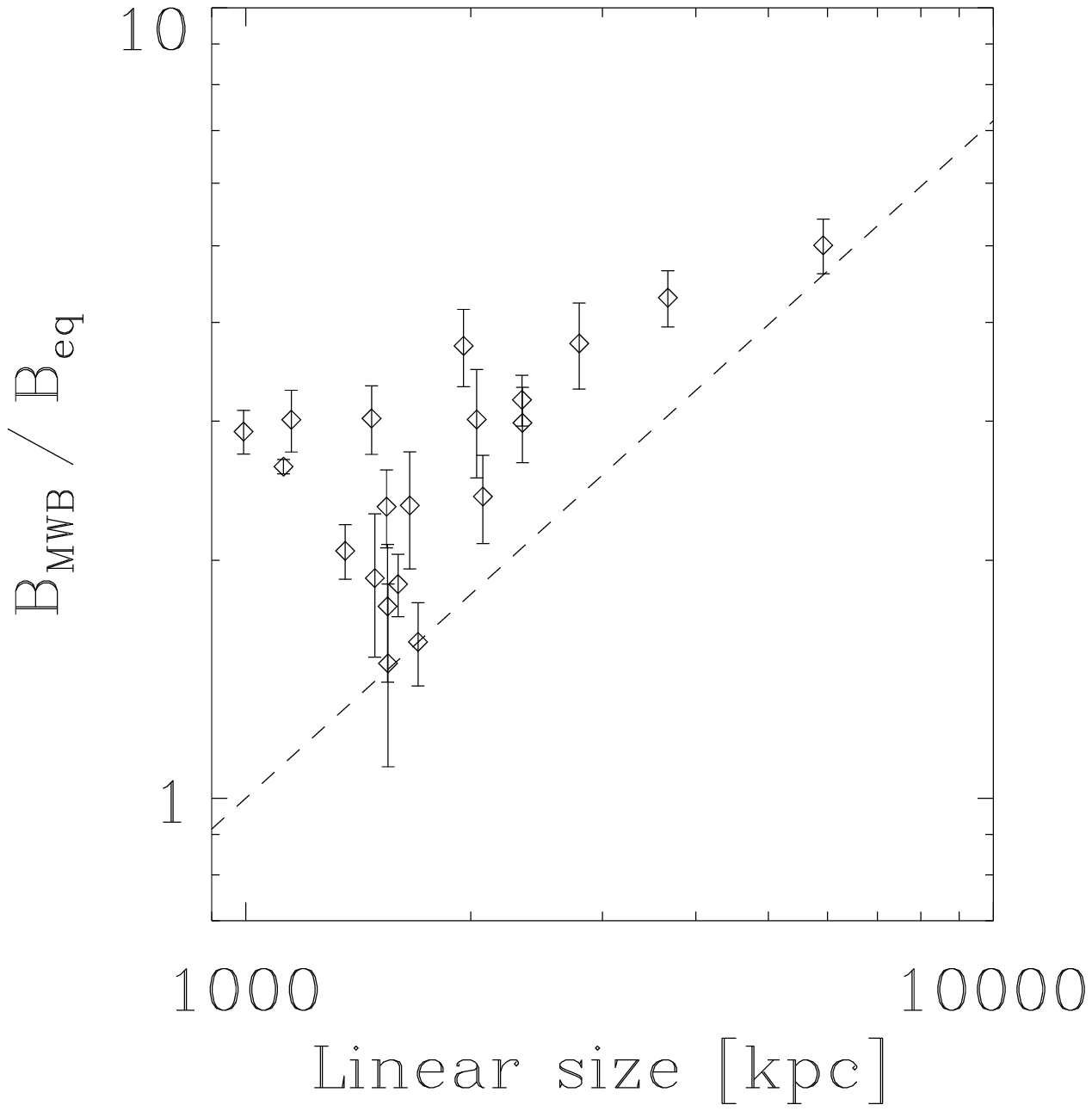}} \\
\end{tabular}
\caption{\label{fig:b_ic-b_eq} The ratio of the equivalent magnetic
field strength of the microwave background radiation, $B_{MWB}$, to
the averaged equipartition magnetic field strengths, $B_{eq}$, in the
lobes of the GRGs against redshift (left) and projected
linear size (right) of the radio source. The dashed line in the right
plot indicates the expected behaviour of $B_{MWB}/B_{eq}$ for increasing
linear size of a source with constant radio power and redshift
($B_{eq} \propto V^{-2/7} \propto D^{-6/7}$, where $V$ and $D$ are the
volume and linear size of the radio source, respectively). This line only
illustrates that the apparent correlation in this diagram is due to selection effects.}
\end{figure*}

We have fitted advance velocities and ages of the radio lobes to the
spectral index profiles
using the method described in Schoenmakers et al. (1998). In short,
this method works as follows: First, we recognize that Inverse Compton
scattering of the Microwave Background (MWB) radiation is an important
energy loss mechanism of the radiating particles in the lobes of
GRGs. Its influence on the energy losses can be described 
by imposing an additional magnetic field,
$B_{MWB}$, whose energy density equals that of the MWB radiation
field, i.e. $B_{MWB} = 3.24\,(1+z)^2 \mu$G, where $z$ is the redshift of
the source.  In Fig. \ref{fig:b_ic-b_eq} we have
plotted the ratio of the equivalent magnetic field strength of the MWB
radiation, $B_{MWB}$, and the equipartition magnetic field strengths
in the radio lobes, $B_{eq}$, averaged for both lobes, against
both redshift and linear size of the GRGs (see also Ishwara-Chandra \&
Saikia 1999). See Sect. \ref{sec:energy_densities} for the calculation
of the equipartition magnetic field strength.  We find that in all GRGs
$B_{eq} \la B_{MWB}$, which
indicates that the IC-scattering process dominates the energy losses
of the radio lobes.  If we neglect energy losses resulting from the expansion
of the lobes, then the radiating particles loose energy only due to synchrotron
radiation and IC scattering of MWB photons. In this case, the
radiative lifetime of particles is maximized when the internal magnetic
field strength of the lobe
$B_{int} = B_{MWB}/\sqrt{3}$ (e.g. van der Laan \& Perola 1969).In
other words, for a given time since the last acceleration of the radiating particles, this magnetic field strength gives the least
amount of spectral steepening due to radiation and IC scattering.  Using
this value for the internal magnetic field strength therefore
provides an upper limit to the spectral age of a radio source.

We have derived the velocities and ages of the radio lobes using
both the equipartition magnetic field strength (see
Tab. \ref{tab:energy-densities}) and the magnetic field strength
which gives the maximum age ($B_{MWB}/\sqrt{3}$).  
We have calculated the velocities and ages
for two ageing models: The Jaffe-Perola (JP) and the Continuous Injection (CI) models (see Schoenmakers et al. 1998, and references therein,
for details). These are the extreme cases in the sense that the JP model
gives the largest change in spectral index for a given age, whereas
the CI model gives the smallest change. We find that in many cases our
data are not good enough to decide which model is better
applicable. It can be expected, however, that this is the JP model
unless reacceleration of the particles in the lobes is important.
Last, we assume that the advance velocity of the radio lobes, the
injection spectral index of the radiating particles and the magnetic
field strength in the lobes have been constant during the life-time
of the radio source, and that the magnetic field strength is uniform
throughout the lobe.

The velocity we find with our method is the separation velocity between
the head of 
the lobe and the material flowing back in the lobe. This backflow
may be important in very powerful radio sources (e.g. Liu et
al. 1992, Scheuer 1995), but whether this is also the case in GRGs is
not clear. Some of our sources have radio lobes which do not cover the
whole area between the hotspot and the radio core. 
Although this may be caused by the increased effect of
spectral ageing at a large distance from the hotspot, it may also
indicate that backflows are not important in these radio
lobes. Although the issue is far from clear, we will assume for the
remainder of this discussion that backflows are
unimportant in the lobes of the GRGs.

\begin{figure*}[t!]
\setlength{\tabcolsep}{0pt}
\begin{tabular}{l l l l}
\resizebox{0.25\hsize}{!}{\epsfig{file=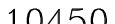,angle=90}} & 
\resizebox{0.25\hsize}{!}{\epsfig{file=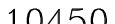,angle=90}} & 
\resizebox{0.25\hsize}{!}{\epsfig{file=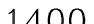,angle=90}} & 
\resizebox{0.25\hsize}{!}{\epsfig{file=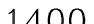,angle=90}} \\
\resizebox{0.25\hsize}{!}{\epsfig{file=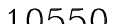,angle=90}} & 
\resizebox{0.25\hsize}{!}{\epsfig{file=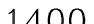,angle=90}} & 
\resizebox{0.25\hsize}{!}{\epsfig{file=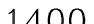,angle=90}} & 
\resizebox{0.25\hsize}{!}{\epsfig{file=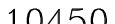,angle=90}} \\ 
\resizebox{0.25\hsize}{!}{\epsfig{file=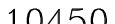,angle=90}} & 
\resizebox{0.25\hsize}{!}{\epsfig{file=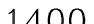,angle=90}} &
\resizebox{0.25\hsize}{!}{\epsfig{file=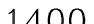,angle=90}} & 
\resizebox{0.25\hsize}{!}{\epsfig{file=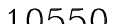,angle=90}} \\ 
\resizebox{0.25\hsize}{!}{\epsfig{file=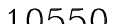,angle=90}} &
\resizebox{0.25\hsize}{!}{\epsfig{file=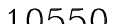,angle=90}} \\
\end{tabular}
\caption{\label{fig:vel-plots} The results of the spectral ageing
analyses. The plotted points are the spectral index points used for
the fit in each lobe of the seven sources and the lines are the best
fits for the two ageing models.  The dotted line represents the CI
model and the dashed line the JP model. The numbers along lower axes denote the
distance in kpc from the point where we assume that the age is zero.  
The numbers along the top axis indicate the distance from the
host galaxy in kpc. The results plotted here are those using an
internal magnetic field strength of $B_{MWB}/\sqrt{3}$, which should
yield the maximum age. For the source B\,1003$+$351 only the western
lobe could be used for fitting. For the southern lobe of B\,2043$+$749 we present two fits, one using seven and one using five spectral index
points, since these give quite different results (see
Tab. \ref{tab:lobe-velocities}).}
\end{figure*}

Many of our sources either do not show the expected ageing
signature, i.e. a steepening of the spectrum towards the radio core,
or have too few (less than five) or too poorly determined datapoints to
provide a meaningful constraint for the fitting process (see
Figs. \ref{fig:si-profiles} and \ref{fig:si-profiles2}).
Only seven sources have spectral index profiles that we could use to
fit the velocities and ages. 
These are B\,0109$+$492, B\,0813$+$758 (between 354
and 1400 MHz), B\,1003$+$351 (western lobe only), B\,1209$+$745
(between 354 and 1400 MHz), B\,1312$+$698, B\,1543$+$845 and B\,2043$+$749. 
 The results
for these are presented in Tab. \ref{tab:lobe-velocities}, and the
model fits in Fig. \ref{fig:vel-plots}.  The plots in
Fig. \ref{fig:vel-plots} have been made for the case that $B_{int} =
B_{MWB}/\sqrt{3}$. In the table we present for each lobe the best fit
velocity, $v$, and injection spectral index, $\alpha_{inj}$, and the
reduced $\chi^2$ of the fit. In case of the source B\,2043$+$749 we are able
to fit the spectral index profile of the southern lobe using either five
or seven spectral index points.  The results are significantly different and
we therefore present them for both these cases. 
The values of the reduced $\chi^2$ presented in
Tab. \ref{tab:lobe-velocities} are almost all smaller than unity. We
believe that this is largely due to our conservative error estimates in the
spectral indices.

We find that the advance velocities of
the heads of the lobes of the GRG are in the range of $0.01c - 0.1c$,
with an average value of $\sim\!0.04c$.
The many sources that could not be fitted using our method clearly question the
validity of our method and the results. If some of the mechanisms that
prevent the presence of a clear
ageing signature in all these sources are also at work in the
seven sources that we were able to find the age for, then the 
velocities that we have obtained are only upper limits to
the true velocities. They indicate a general trend, though, that the
advance velocities of the lobes of GRGs are below $0.1c$. 
  
\begin{table*}[t!]
\caption{\label{tab:lobe-velocities} Results of the spectral aging
analysis.  The top part of the table presents the results for an
internal magnetic field strength equal to $B_{MWB}/\sqrt{3}$, which
should give the maximum age; the bottom part is for an internal field
strength equal to the average equipartition field strength in the two
lobes. Column 1 gives the name of the source and the component. Column
2 gives the number of spectral index points used in the
fitting. Columns 3 to 5 give the velocity, $v$, in units of $c$, the
injection spectral index, $\alpha_{inj}$ and the reduced $\chi^2$ of
the fit using the CI model. Columns 6 to 8 give the same for the JP
model. Column 9 and 10 give the age of the lobes resulting from the
fitted velocities and the size of the lobes, for each of the two
models.}
\begin{tabular}{l c c r@{$\,\pm\,$}r r@{$\,\pm\,$}r r r@{$\,\pm\,$}r r@{$\,\pm\,$}r r r@{$\,\pm\,$}r r@{$\,\pm\,$}r}
\hline \hline \\
\multicolumn{2}{c}{(1)} & \multicolumn{1}{c}{(2)} & \multicolumn{2}{c}{(3)} & \multicolumn{2}{c}{(4)} & \multicolumn{1}{c}{(5)} & \multicolumn{2}{c}{(6)} & \multicolumn{2}{c}{(7)} & \multicolumn{1}{c}{(8)} &
 \multicolumn{2}{c}{(9)} & \multicolumn{2}{c}{(10)} \\[1ex]
\multicolumn{2}{c}{Source} & \multicolumn{1}{c}{} & \multicolumn{2}{c}{$v_{\rm CI}$} & \multicolumn{2}{c}{$\alpha_{inj, \rm CI}$}  & \multicolumn{1}{c}{$\chi^2_{red}$} & \multicolumn{2}{c}{$v_{\rm JP}$} & \multicolumn{2}{c}{$\alpha_{inj, \rm JP}$} & \multicolumn{1}{c}{$\chi^2_{red}$} & \multicolumn{2}{c}{$t_{\rm CI}$} & \multicolumn{2}{c}{$t_{\rm JP}$}\\
 & & & \multicolumn{2}{c}{$[\,c\,]$} & \multicolumn{2}{c}{\ } & & \multicolumn{2}{c}{$[\,c\,]$} &  \multicolumn{2}{c}{\ } &  & \multicolumn{4}{c}{$[$\,$10^7$~yr\,$]$} \\[1ex]
\hline \\
B\,0109$+$492 & N &  6 & 0.014 & 0.001 & $-0.72$ & 0.01 & 0.38 & 0.040 & 0.001 & $-0.74$ & 0.01 & 0.46 & 13.1 & 0.9 &  4.6 & 0.1 \\
            & S &  5 & 0.014 & 0.001 & $-0.69$ & 0.01 & 0.68 & 0.029 & 0.001 & $-0.69$ & 0.01 & 0.86 & 13.1 & 0.9 &  6.2 & 0.2 \\[1ex]
B\,0813$+$758 & E & 11 & 0.046 & 0.001 & $-0.71$ & 0.03 & 0.49 & 0.094 & 0.001 & $-0.71$ & 0.03 & 0.34 & 10.5 & 1.3 &  5.1 & 0.7 \\
            & W &  6 & 0.032 & 0.001 & $-0.81$ & 0.04 & 0.55 & 0.061 & 0.001 & $-0.81$ & 0.04 & 0.43 &  8.8 & 0.3 &  4.6 & 0.1 \\[1ex]
B\,1003$+$351 & W &  8 & 0.059 & 0.002 & $-0.70$ & 0.02 & 0.09 & 0.119 & 0.002 & $-0.70$ & 0.02 & 0.06 & 13.1 & 0.4 &  6.5 & 0.1 \\[1ex]
B\,1209$+$745 & N &  6 & 0.017 & 0.001 & $-0.81$ & 0.03 & 0.07 & 0.033 & 0.001 & $-0.81$ & 0.03 & 0.09 & 14.0 & 1.3 &  7.2 & 0.6 \\
            & S &  6 & 0.011 & 0.001 & $-0.78$ & 0.03 & 0.34 & 0.024 & 0.001 & $-0.79$ & 0.03 & 0.47 & 12.5 & 1.2 &  5.7 & 0.3 \\[1ex]
B\,1312$+$698 & E &  6 & 0.037 & 0.002 & $-0.71$ & 0.02 & 0.42 & 0.103 & 0.002 & $-0.74$ & 0.02 & 0.17 &  8.0 & 0.4 &  2.9 & 0.1 \\
            & W &  5 & 0.024 & 0.002 & $-0.75$ & 0.02 & 0.96 & 0.056 & 0.002 & $-0.76$ & 0.02 & 0.57 &  6.1 & 0.5 &  2.6 & 0.1 \\[1ex]
B\,1543$+$845 & N &  7 & 0.025 & 0.001 & $-0.75$ & 0.04 & 1.17 & 0.051 & 0.001 & $-0.74$ & 0.04 & 0.58 & 12.7 & 0.7 &  6.2 & 0.2 \\
            & S &  8 & 0.021 & 0.001 & $-0.65$ & 0.04 & 0.96 & 0.045 & 0.001 & $-0.65$ & 0.03 & 0.34 & 17.2 & 1.3 &  8.0 & 0.3 \\[1ex] 
B\,2043$+$749 & N &  5 & 0.029 & 0.001 & $-0.70$ & 0.02 & 0.27 & 0.057 & 0.001 & $-0.70$ & 0.02 & 0.07 & 8.6  & 0.3 &  4.3 & 0.1 \\         
            & S &  5 & 0.030 & 0.002 & $-0.77$ & 0.02 & 0.52 & 0.060 & 0.002 & $-0.77$ & 0.02 & 0.42 & 8.5  & 0.5 &  4.3 & 0.2 \\
            & S &  7 & 0.059 & 0.002 & $-0.74$ & 0.01 & 0.83 & 0.112 & 0.002 & $-0.74$ & 0.01 & 0.55 &  4.3 & 0.2 &  2.3 & 0.1 \\
\hline \\
B\,0109$+$492 & N &  6 & 0.016 & 0.001 & $-0.72$ & 0.01 & 0.39 & 0.045 & 0.001 & $-0.73$ & 0.01 & 0.51 & 11.4 & 0.7 &  4.1 & 0.1 \\
            & S &  5 & 0.016 & 0.001 & $-0.70$ & 0.02 & 0.73 & 0.033 & 0.001 & $-0.70$ & 0.01 & 0.24 & 11.4 & 0.7 &  5.5 & 0.2 \\[1ex]
B\,0813$+$758 & E & 11 & 0.054 & 0.002 & $-0.71$ & 0.03 & 0.48 & 0.110 & 0.002 & $-0.71$ & 0.03 & 0.34 &  8.9 & 1.2 &  4.4 & 0.5 \\
            & W &  6 & 0.037 & 0.001 & $-0.81$ & 0.03 & 0.55 & 0.071 & 0.001 & $-0.81$ & 0.04 & 0.44 &  7.7 & 0.4 &  4.0 & 0.2 \\[1ex]
B\,1003$+$351 & W &  8 & 0.091 & 0.002 & $-0.69$ & 0.02 & 0.13 & 0.173 & 0.002 & $-0.70$ & 0.02 & 0.06 &  8.5 & 0.2 &  4.5 & 0.1 \\[1ex] 
B\,1209$+$745 & N &  6 & 0.019 & 0.001 & $-0.81$ & 0.03 & 0.08 & 0.038 & 0.001 & $-0.81$ & 0.03 & 0.09 & 12.5 & 1.1 &  6.2 & 0.5 \\
            & S &  6 & 0.013 & 0.001 & $-0.78$ & 0.03 & 0.37 & 0.028 & 0.001 & $-0.79$ & 0.02 & 0.48 & 10.6 & 0.9 &  4.9 & 0.2 \\[1ex]
B\,1312$+$698 & E &  6 & 0.038 & 0.002 & $-0.71$ & 0.02 & 0.41 & 0.105 & 0.002 & $-0.74$ & 0.02 & 0.25 &  7.8 & 0.4 &  2.8 & 0.1 \\
            & W &  5 & 0.025 & 0.002 & $-0.75$ & 0.02 & 0.97 & 0.055 & 0.002 & $-0.75$ & 0.02 & 0.63 &  5.8 & 0.5 &  2.7 & 0.1 \\[1ex]
B\,1543$+$845 & N &  7 & 0.027 & 0.001 & $-0.75$ & 0.03 & 1.17 & 0.055 & 0.001 & $-0.74$ & 0.03 & 0.58 & 11.8 & 0.7 &  5.8 & 0.2 \\
            & S &  8 & 0.022 & 0.001 & $-0.65$ & 0.04 & 0.96 & 0.049 & 0.001 & $-0.65$ & 0.03 & 0.34 & 16.4 & 1.3 &  7.3 & 0.3 \\[1ex] 
B\,2043$+$749 & N &  5 & 0.034 & 0.001 & $-0.70$ & 0.02 & 0.34 & 0.067 & 0.001 & $-0.70$ & 0.02 & 0.15 & 7.3  & 0.2 &  3.7 & 0.1 \\         
            & S &  5 & 0.036 & 0.002 & $-0.76$ & 0.02 & 0.63 & 0.071 & 0.002 & $-0.76$ & 0.02 & 0.51 & 7.1  & 0.4 &  3.6 & 0.2 \\
            & S &  7 & 0.047 & 0.002 & $-0.71$ & 0.01 & 0.77 & 0.093 & 0.002 & $-0.71$ & 0.01 & 0.64 &  5.4 & 0.3 &  2.7 & 0.2 \\
\hline \hline \\
\end{tabular}
\end{table*}

\subsection{The spectral ages of GRGs}

The method to find the
velocities of the lobes which we employed in Sect. \ref{sec:velocities}
also yields the spectral age of the radio source. Using the lobe
advance velocities in Tab. \ref{tab:lobe-velocities} and the known armlengths
of the lobes (Tab. \ref{tab:morphologies}), we have calculated the lobe
ages.  We have calculated the ages for each of the two spectral ageing models and for both the
equipartition magnetic field strength and the one that should maximize
the age of the lobe, $B_{MWB}/\sqrt{3}$. 

In Tab. \ref{tab:lobe-velocities} we present the spectral ages of the
lobes of the seven GRGs in our sample for which we could find the lobe
velocities. In general, the differences in the spectral age for 
the two values of the magnetic field strength are
small. In reality the age of the two lobes of a single source should be equal.
In all sources where we could fit both lobes, we indeed find that the
ages are close to each other.  In the southern lobe of the source
B\,2043$+$749, which we have fitted using both five and seven spectral
index points, the best agreement between the ages of the two lobes is
found for the fit which uses five spectral points only.

The spectral ages we find all lie in the range between 30 and 150 Myr,
depending on the spectral ageing model and, to a lesser degree, on the
used value of the internal magnetic fieldstrength.  Also, they are
comparable to those found for other Giant radio sources, using similar
techniques, e.g. 40 Myr for B\,0136$+$396 (Hine 1979) and
B\,0821$+$695 (Lacy et al. 1993), 180 Myr in B\,0319$-$454 (Saripalli
et al. 1994), 140 Myr in B\,0313$+$683 (Schoenmakers et al. 1998). We
find an average age of 80 Myr.

\section{The energy densities and pressures of the radio lobes}
\label{sec:energy_densities}

The pressure in the bridges of radio lobes of FRII-type sources are 
higher than that of their environment (e.g. Subrahmanyan \&
Saripalli 1993, Subrahmanyan et al. 1996). Radio source evolution
models predict that the pressure in the lobes decreases with increasing
source size (e.g. Kaiser \& Alexander 1997), so that the lobes of large radio
sources should be closer to pressure equilibrium with their
environment. Since GRGs are the largest radio sources known, they are
best suited to constrain the pressure in the ambient medium. In this
section we investigate the energy densities in the lobe of the
FRII-type GRGs in our sample and we relate this to the properties of
their environment, the IGM.

\subsection{Energy densities of the radio lobes}
\label{sec:en-profiles}

We have calculated the equipartition energy densities and magnetic
field strengths in the lobes of the FRII-type GRGs in our sample. We
have used a method similar to that outlined by Miley (1980). To
determine the luminosity of each lobe, we have used the integrated
325-MHz flux densities (see Tab. \ref{tab:lobe-fluxes}).
The spectral index of each lobe has been calculated between 325 MHz
and 10.5 GHz. In case of significant steepening of the radio spectrum 
at frequencies below
10.5 GHz this value will be too low, but the influence of this on the
result is only marginal (see Miley 1980).

The volumes of the radio lobes have been estimated assuming a
cylindrical morphology. The width has been taken as the deconvolved
average full width of the lobe, measured between the 3$\sigma$
contours on a radio contour map.  The length of the lobe, whenever a
clear gap exists between the core and the tail of the radio lobe (see, e.g.,
B\,0648$+$733, B\,2147$+$816), has been taken as the distance between the
innermost and outermost edge of the lobe at 325 MHz. The radio axes
have been assumed to be in the plane of the sky; although this is
certainly not the case for the majority of sources, the correction
factors are not large compared to the uncertainties introduced by
assuming cylindrical morphologies. 
These used lobe dimensions are 
quoted in Tab. \ref{tab:energy-densities}.  Further assumptions
are a filling factor of the radiating particles of unity and an equal
distribution of energy between electrons/positrons and heavy
particles such as protons. The low and high-frequency cut-offs in the
radio spectra have been set at 10~MHz and 100~GHz, respectively.

In Tab. \ref{tab:energy-densities} we present the equipartition
magnetic field strengths and energy densities of the lobes of the
FRII-type sources in our sample. We find that the magnetic field
strengths are low, typically a few $\mu$G. Consequently, 
also the energy densities, which are directly derived from the
magnetic field strengths ($u \propto B^2$), are low. 
This is as expected since the GRGs are extremely large and not exceptionally powerful.

\begin{table*}[tb]
\setlength{\tabcolsep}{5pt}
\caption{\label{tab:energy-densities}
The equipartition magnetic field strengths and energy densities of the
two lobes of the FRII-type sources in the 1-Jy sample. We have omitted
the source 0309+411 since it is strongly core dominated which did not
allow an accurate measurement.  Column 1 gives the name of the
source. Column 2 gives the sidedness indicator of the lobes `$A$' and
`$B$', where `N' stands for north, `E' for east, etc. Columns 3 to 6
give the lengths, $l$, and widths, $w$, of the lobes (note, that $l$
only gives the part of the lobe from which radio emission has been
detected).  Column 7 gives the reference for the observations we used
to determine $l$ and $w$.  Columns 8 and 9 give the equipartition
magnetic field strength and columns 10 and 11 give the equipartition
energy densities.}
\begin{tabular}{l c r@{$\,\pm\,$}r r@{$\,\pm\,$}r r@{$\,\pm\,$}r r@{$\,\pm\,$}r c r@{$\,\pm\,$}r
r@{$\,\pm\,$}r r@{$\,\pm\,$}r r@{$\,\pm\,$}r}
\hline \hline\\
\multicolumn{1}{c}{(1)} & \multicolumn{1}{c}{(2)} & \multicolumn{2}{c}{(3)} & \multicolumn{2}{c}{(4)} & \multicolumn{2}{c}{(5)} & \multicolumn{2}{c}{(6)} & \multicolumn{1}{c}{(7)} & \multicolumn{2}{c}{(8)} &
 \multicolumn{2}{c}{(9)} & \multicolumn{2}{c}{(10)} & \multicolumn{2}{c}{(11)}\\
Source & \multicolumn{1}{c}{A,B} & \multicolumn{2}{c}{$l_A$} & \multicolumn{2}{c}{$w_A$}  & \multicolumn{2}{c}{$l_B$} & \multicolumn{2}{c}{$w_B$} & Ref. & \multicolumn{2}{c}{$B_{eq,A}$} & \multicolumn{2}{c}{$B_{eq,B}$} & \multicolumn{2}{c}{$u_{eq,A}$} & \multicolumn{2}{c}{$u_{eq,B}$} \\
 & & \multicolumn{2}{c}{$[$\,\arcsec\,$]$} & \multicolumn{2}{c}{$[$\,\arcsec\,$]$} & \multicolumn{2}{c}{$[$\,\arcsec\,$]$} &  \multicolumn{2}{c}{$[$\,\arcsec\,$]$} &  & \multicolumn{4}{c}{$[$\,$\mu$G\,$]$} &  \multicolumn{4}{c}{$[$\,$10^{-14}$ erg cm$^{-3}$\,$]$} \\
\hline \\
B\,0050$+$402 & N,S &  120 &  10 &   30 &  10 &  160 &  10 &   40 &  10 & \phantom{1}1 & 3.05 & 0.93 & 2.72 & 0.57 & 86.3 & 37.3 & 68.7 & 20.2 \\
B\,0109$+$492 & N,S &  323 &   2 &  210 &   5 &  323 &   2 &  210 &   5 & \phantom{1}2 & 1.44 & 0.03 & 1.36 & 0.03 & 19.3 &  0.6 & 17.3 &  0.5 \\
B\,0136$+$396 & E,W &  201 &   8 &   90 &  10 &  164 &   8 &   90 &  10 & \phantom{1}3 & 1.81 & 0.16 & 3.28 & 0.30 & 30.5 &  3.9 & 99.7 & 12.8 \\
B\,0211$+$326 & E,W &  167 &   5 &   40 &  10 &  140 &   5 &   40 &  10 & \phantom{1}1 & 2.85 & 0.56 & 3.03 & 0.60 & 75.6 & 21.0 & 85.3 & 23.9 \\
B\,0648$+$733 & E,W &  200 &  30 &  130 &  20 &  200 &  20 &  130 &  20 & \phantom{1}4 & 0.98 & 0.18 & 1.69 & 0.24 &  9.0 &  2.3 & 26.4 &  5.3 \\
B\,0813$+$758 & E,W &  250 &  40 &  100 &  10 &  140 &  10 &  100 &  10 & \phantom{1}4 & 1.74 & 0.21 & 1.56 & 0.15 & 28.0 &  4.8 & 22.6 &  3.0 \\
B\,0945$+$734 & E,W &  370 &   3 &  385 &  10 &  440 &  10 &  250 &  50 & \phantom{1}2 & 0.96 & 0.02 & 1.43 & 0.22 &  8.6 &  0.3 & 19.1 &  4.0 \\
B\,1003$+$351 & E,W & 1450 &  10 &  160 &  20 &  980 &  10 &  340 &  30 & \phantom{1}5 & 0.86 & 0.08 & 0.70 & 0.04 &  6.9 &  0.9 &  4.6 &  0.4 \\
B\,1209$+$745 & N,S &  150 &  30 &  170 &  10 &  112 &  10 &  180 &  10 & \phantom{1}2 & 1.29 & 0.14 & 1.35 & 0.09 & 15.3 &  2.4 & 16.9 &  1.6 \\
B\,1213$+$422 & N,S &  168 &   3 &   40 &  10 &  141 &  10 &   40 &  10 & \phantom{1}6 & 2.57 & 0.50 & 2.70 & 0.58 & 61.2 & 16.9 & 67.7 & 20.5 \\
B\,1309$+$412 & N,S &  186 &   3 &  120 &  10 &  186 &   3 &  120 &  10 & \phantom{1}3 & 1.44 & 0.09 & 1.31 & 0.08 & 19.2 &  1.7 & 15.8 &  1.4 \\
B\,1312$+$698 & E,W &  352 &   5 &   95 &  10 &  174 &   5 &   95 &  10 & \phantom{1}7 & 1.53 & 0.12 & 2.32 & 0.19 & 21.8 &  2.3 & 50.1 &  5.7 \\
B\,1358$+$305 & N,S &  200 &   8 &  210 &  20 &  180 &  30 &  130 &  20 & \phantom{1}8 & 1.26 & 0.10 & 1.24 & 0.21 & 14.7 &  1.6 & 14.4 &  3.5 \\
B\,1426$+$295 & E,W &  365 &  10 &  140 &  20 &  350 &  20 &  115 &  15 & \phantom{1}9 & 0.99 & 0.11 & 1.06 & 0.12 &  9.1 &  1.4 & 10.4 &  1.7 \\
B\,1450$+$333 & N,S &  100 &  10 &   40 &   5 &  120 &  10 &   40 &   5 & \phantom{1}6 & 3.71 & 0.45 & 2.70 & 0.32 &127.5 & 21.8 & 67.8 & 11.5 \\
B\,1543$+$845 & N,S &  170 &  20 &   60 &   5 &  185 &  20 &   60 &  10 & \phantom{1}4 & 2.03 & 0.20 & 1.85 & 0.29 & 38.3 &  5.4 & 31.7 &  7.1 \\
B\,1626$+$518 & N,S &  360 &  30 &   45 &  10 &  320 &  30 &  150 &  20 & 10& 1.84 & 0.36 & 1.24 & 0.15 & 31.3 &  8.8 & 14.2 &  2.5 \\
B\,1918$+$516 & N,S &  180 &   5 &   75 &   5 &  210 &  10 &   90 &  10 & \phantom{1}7 & 2.00 & 0.11 & 1.35 & 0.14 & 37.1 &  2.9 & 16.8 &  2.4 \\
B\,2043$+$749 & N,S &  195 &  20 &  115 &  15 &  195 &  20 &  150 &  15 & 11& 1.62 & 0.18 & 1.80 & 0.17 & 24.2 &  4.0 & 30.2 &  4.0 \\
B\,2147$+$816 & N,S &  370 &  20 &  120 &  10 &  310 &  20 &  120 &  10 & \phantom{1}9 & 0.94 & 0.08 & 1.04 & 0.09 &  8.1 &  0.9 & 10.1 &  1.2 \\
\hline \hline\\
\end{tabular}
\begin{minipage}{\linewidth}
References: (1) 1.4-GHz WSRT data (paper I); (2) J\"{a}gers 1986; (3) Vigotti et al. 1989; (4) Lara et al. (in preparation); (5) Mack et al. 1997; (6) FIRST survey; (7) 1.4-GHz WSRT data (unpublished); (8) Parma et al. 1996; (9) NVSS survey; (10) R\"{o}ttgering et al. 1996; (11) Riley \& Warner 1990.\\
\end{minipage}
\end{table*} 

\begin{figure*}[tb]
\setlength{\tabcolsep}{0pt}
\begin{tabular}{l l l}
\resizebox{0.33\hsize}{!}{\epsfig{file=0109+492.u_eqplot.eps,angle=90}} & 
\resizebox{0.33\hsize}{!}{\epsfig{file=0813+758.u_eqplot.eps,angle=90}} &
\resizebox{0.33\hsize}{!}{\epsfig{file=0945+734.u_eqplot.eps,angle=90}} \\ 
\resizebox{0.33\hsize}{!}{\epsfig{file=1003+351.u_eqplot.eps,angle=90}} & 
\resizebox{0.33\hsize}{!}{\epsfig{file=1209+745.u_eqplot.eps,angle=90}} & 
\resizebox{0.33\hsize}{!}{\epsfig{file=1312+698.u_eqplot.eps,angle=90}} \\
\resizebox{0.33\hsize}{!}{\epsfig{file=1358+305.u_eqplot.eps,angle=90}} &
\resizebox{0.33\hsize}{!}{\epsfig{file=1426+295.u_eqplot.eps,angle=90}} & 
\resizebox{0.33\hsize}{!}{\epsfig{file=1543+845.u_eqplot.eps,angle=90}} \\ 
\resizebox{0.33\hsize}{!}{\epsfig{file=1626+518.u_eqplot.eps,angle=90}} &
\resizebox{0.33\hsize}{!}{\epsfig{file=2043+749.u_eqplot.eps,angle=90}} &
\resizebox{0.33\hsize}{!}{\epsfig{file=2147+816.u_eqplot.eps,angle=90}}  \\
\end{tabular}
\caption{\label{fig:ueq-profiles} Profiles of the equipartition energy
density distribution along the major axes of several of the radio sources.
The zeropoint is the position of the radio core. The numbers along the
upper axis denote the distance from the core in units of kpc. Negative
values are on that side of the source which is mentioned first in
column two of Tab. \ref{tab:energy-densities}.}
\end{figure*}

\subsection{Pressure gradients in the lobes}
\label{sec:pressure-gradients}

The equipartition energy densities in Tab. \ref{tab:energy-densities}
are the lobe volume averaged energy densities. Within the lobes,
however, large variations may be present. To investigate this, we have made
profiles of the equipartition energy density for the FRII-type sources
above $7\arcmin$ in size. We have used the same method as we have used
to find the spectral index profiles in Sect. \ref{sec:si-profiles}, i.e
by superposing on the source an array of boxes of width half the FWHM
beam-size. In each box we have determined the average deconvolved
width of the lobe by measuring the distance between the $3\sigma$ outer
contours, and we have integrated the 325/354-MHz flux density. The
spectral index of each box was calculated between 325/354 MHz
and 10.5 GHz. In the case of B\,0813$+$758 and B\,1209$+$745 we used
the 1.4-GHz NVSS map as the high-frequency map in order to employ
the higher resolution available. In all cases the radio maps
were first convolved to the lowest resolution map
used.  Only in boxes where the integrated flux density at both frequencies
exceeded $3\sigma$, the spectral index was calculated and the
equipartition energy density determined. In case of the sources
B\,0050$+$402 and B\,0648$+$733, the deconvolved lobe widths were below
zero, and we have therefore omitted these sources from further analysis.
The reason for this is not entirely clear to us.

The profiles of the equipartition energy density for the remaining
sources are presented in
Fig. \ref{fig:ueq-profiles}. In some cases, such as 
B\,0813$+$758, B\,0945$+$734 and B\,1312$+$698, we find the largest
values of $u_{eq}$ at 
the heads of the lobes. In some other cases, such as B\,1209$+$745 and
B\,2043$+$749, we clearly detect the influence of the radio core. When we
compare the lobe averaged energy densities as presented in Tab. \ref{tab:energy-densities} with
the average value of the profiles in the plots, 
we find that the differences, as
expected, are relatively small. In only a few cases we find
discrepancies, exceeding a
factor of two, and these are mostly due to the difference in the
assumed geometry of the lobes.

In a relativistic plasma, such as which constitutes the radio lobes, 
the pressure, $p$, is directly related to the
energy density by $p = \frac{1}{3}u_{eq}$. Therefore, the plots in
Fig. \ref{fig:ueq-profiles} also show the behaviour of the lobe
pressure as a function of position along the radio axis.  Since we
have used a relatively low frequency for the flux density measurements
and we have measured the spectral index for each plotted point
separately, the results should not be highly sensitive to the increased effect
of spectral ageing of the electrons towards the radio core.

If the host galaxy is at the center of a cluster, or if it has an
extensive gaseous halo, hydrostatic pressure equilibrium requires a
radial decrease of the pressure. The gradient should be
strongest at small radii, around the core radius of the gas
distribution. If the pressures in the bridges of the lobes are close
to pressure equilibrium with their surrounding, it is expected that
this gradient should be reflected in the lobe pressures at small
radii.

We do not find such a behaviour in the majority of our sources. In the few cases where
the energy density appears to increase near the position of the host
galaxy this can be attributed to the presence of a radio core and/or radio jets
(e.g. B\,1209$+$745 and B\,2043$+$749). Only in B\,1543$+$845 and in the
eastern lobe of B\,0813$+$758 we observe a small increase in the
pressure at small radii which cannot be related to a
strong radio core or a jet (L. Lara, priv. comm.), and which might
thus indicate the presence of a pressure gradient in the environment.
 
In most sources the energy density actually increases with increasing
distance from the host galaxy. This indicates that the lobes must be
overpressured with respect to their environment, even at small radii,
and that there are pressure gradients in these lobes: The heads
of the lobes have much higher pressures than the bridges. 

\subsection{The density of the environment of the lobes}

From the ages, advance velocities and energy densities of the lobes we
can estimate the density of the environment. We assume that the
propagation of the head of the lobe is governed by a balance between
the thrust of the jet and the ram-presssure exerted by the
environment. In this case, the external density, $\rho_a$, is given by
$\rho_a = \Pi_j / (A_h v_h^2)$, where $\Pi_j$ is the thrust of
the jet, $A_h$ is the area of the bowshock and $v_h$ is the advance
velocity of the head of the lobe.  The thrust $\Pi_j$ is given by
$Q_{jet} / v_j$, with $v_j$ the velocity of the material in the
jet, which we assume to be $c$, the velocity of light, and $Q_{jet}$,
which is the amount of energy delivered by the jet per unit time, or
the jet power. We can estimate the jet power by dividing the total
energy contents of the lobes of the radio source by the age of the
source. An important factor in this is the efficiency with which jet
energy is converted to radiation. A conservative choice is given by
Rawlings \& Saunders (1991): Half of the energy goes into the
expansion of the lobe and the other half into radiation. Blundell et
al. (1999), on the basis of radio source modelling, estimate values
between 0.3 and 0.6 which agrees with a value of 0.5\,. We therefore
assume that the jet power is twice as high as the energy content
of a radio lobe, divided by its age.  The energy content can be
estimated by multiplying the equipartition energy density with the
volume of the lobe.

Since $\rho_a \!\propto\! A_h^{-1} \!\propto\! D_h^{-2}$ (where $D_h$
is the diameter of the impact area), $D_h$ plays an important role in
determining the ambient density.  We cannot constrain it directly from
our observations, but a good estimate is given by the size of the
observable hotspot (e.g. Hardcastle et al. 1998).  High-resolution
observations of the hotspots in B\,1312$+$698 can be found in Saunders
et al. (1987). These show that the size of the hotspots is less than
15 kpc. Arcsecond resolution observations of the source B\,2043$+$748
give similar constraints (Riley et al. 1988). Further, Hardcastle et
al. (1998) find a positive correlation between the linear size of a
source and the diameter of the hotspots in a sample of powerful 3CR
FRII-type radio sources at $z<0.3$. This is confirmed for a larger
sample of sources by Jeyakumar \& Saikia (2000). Based on these
results, we have used a diameter of 5 kpc to calculate the ambient
densities of the lobes.  For the age of a source, we have used
the average age of both its lobes, and of both, CI and JP, ageing
models. We have used the ages calculated with an internal magnetic
field strength of $B_{MWB}/\sqrt{3}$, since these are upper limits to
the age and thus provide lower limits to $Q_{jet}$ and $v_h$.  Using
this average age we have calculated the average jet power of the two
lobes, which we assume to be equal on both sides of the source.  
The resulting densities
are presented in Tab. \ref{tab:densities}.  

We find particle densities
between $1\!\times\!10^{-5}$ and $1\!\times\!10^{-4}$ cm$^{-3}$.  They
have been calculated using a mean atomic mass of 1.4 amu in the
environment of the radio sources.  The densities are in good agreement
with the results of Mack et al. (1998).
If we assume a temperature in the IGM of a few $10^6$ K (e.g. Cen \&
Ostriker 1999), the thermal pressure in the environment would be
$\sim\!2\times10^{-14}$ dyn cm$^{-2}$ for a particle density of
$4\times10^{-5}$ cm$^{-3}$.  Such low pressures support the
indications in Sect. \ref{sec:pressure-gradients} that the radio
lobes must be overpressured with respect to the IGM.

\begin{table*}[tb]
\centering
\caption{\label{tab:densities}
Properties of the environment of the seven GRGs with useful lobe
velocity and age determination. Column 1 gives the name of the
source. Column 2 gives the volume, $V$, of the radio source. Column 3
gives the total energy, $u_{tot}$, of the radio lobes, defined as the
equipartition energy density times the volume. Column 4 gives the
average age, $t$, of the lobe. Column 5 gives the jet power,
$Q_{jet}$. Column 6 gives the ambient density, $\rho_a$, of the lobe
and column 7 gives the particle density, $n_a$, assuming a mean
particle mass of 1.4 amu. The densities have been calculated using the
mean of the jet powers of the lobes printed in column 5 and a diameter
of the jet impact area of 5 kpc.}
\begin{tabular}{l c r@{$\,\pm\,$}l r@{$\,\pm\,$}l r@{$\,\pm\,$}l l@{$\,\pm\,$}l l@{$\,\pm\,$}l r@{$\,\pm\,$}l}
\hline \hline \\
\multicolumn{2}{c}{(1)} & \multicolumn{2}{c}{(2)} & \multicolumn{2}{c}{(3)} & \multicolumn{2}{c}{(4)} & \multicolumn{2}{c}{(5)} & \multicolumn{2}{c}{(6)} & \multicolumn{2}{c}{(7)} \\
\multicolumn{2}{c}{Source} & \multicolumn{2}{c}{$V$} & \multicolumn{2}{c}{$u_{tot}$}  & \multicolumn{2}{c}{$t$} & \multicolumn{2}{c}{$Q_{jet}$} & \multicolumn{2}{c}{$\rho_a$} & \multicolumn{2}{c}{$n_a$} \\
 & & \multicolumn{2}{c}{$[\,10^{72}$ cm$^3\,]$} & \multicolumn{2}{c}{$[\,10^{59}$ erg\,$]$} & \multicolumn{2}{c}{$[\,10^7$ yr\,$]$} & \multicolumn{2}{c}{$[\,10^{44}$ erg s$^{-1}\,]$} & \multicolumn{2}{c}{$[\,10^{-29}$ gr cm$^{-3}\,]$} & \multicolumn{2}{c}{$[\,10^{-5}$ cm$^{-3}\,]$}\\[1ex]
\hline \\
B\,0109$+$492 & N & 1.73     & 0.06 & 3.34     & 0.16 &  9.3 & 0.6 & \phantom{1}2.16 & 0.16 & \phantom{11}10.7  & 1.1  & \phantom{1}4.56 & 0.47\\
            & S & 1.73     & 0.06 & 2.99     & 0.13 &  9.3 & 0.6   & \phantom{1}2.16 & 0.16 & \phantom{11}10.7  & 1.1  & \phantom{1}4.56 & 0.47\\
B\,0813$+$758 & E & 5.91     & 1.26 & 16.5 & 4.5  &  7.3 & 0.6     & \phantom{1}10.2 & 2.7  & \phantom{11}4.64 & 1.58 & \phantom{1}1.98 & 0.67\\
            & W & 3.31     & 0.52 &  7.48    & 1.54 &  7.3 & 0.6 &   \phantom{1}10.2 & 2.7  & \phantom{11}13.3  & 3.8  & \phantom{1}5.68 & 1.62\\
B\,1003$+$351 & W & 37.8 & 3.4  & 17.4 & 2.2  &  9.8 & 0.3 &         \phantom{1}10.2 & 1.4  & \phantom{11}2.88 & 0.40 & \phantom{1}1.24 & 0.16\\
B\,1209$+$745 & N & 1.76     & 0.38 &  2.69    & 0.72 &  9.9 & 0.9 & \phantom{1}1.67 & 0.39 & \phantom{11}5.74 & 1.52 & \phantom{1}2.45 & 0.65\\
            & S & 1.48     & 0.18 &  2.50    & 0.39 &  9.9 & 0.9 &   \phantom{1}1.67 & 0.39 & \phantom{11}16.9  & 4.1  & \phantom{1}7.22 & 1.75\\ 
B\,1312$+$698 & E & 1.26     & 0.19 &  2.75    & 0.51 &  5.0 & 0.3 & \phantom{1}3.83 & 0.77 & \phantom{11}2.18 & 0.44 & \phantom{1}9.31 & 0.89\\
            & W & 0.62     & 0.09 &  3.11    & 0.57 &  5.0 & 0.3 &   \phantom{1}3.83 & 0.77 & \phantom{11}9.02 & 1.85 & \phantom{1}3.85 & 0.79\\
B\,1543$+$845 & N & 3.21     & 0.48 & 12.3 & 2.5  & 11.0 & 0.8 &     \phantom{1}9.18 & 2.83 & \phantom{11}21.6  & 6.7  & \phantom{1}9.22 & 2.84\\
            & S & 6.16     & 1.27 & 19.5 & 5.9 & 11.0 & 0.8 &        \phantom{1}9.18 & 2.83 & \phantom{11}16.7 &  5.1  & \phantom{1}7.13 & 2.20\\
B\,2043$+$758 & N & 0.94     & 0.18 &  2.27    & 0.58 &  6.5 & 0.3 & \phantom{1}3.48 & 0.82 & \phantom{11}4.77 & 0.41 & \phantom{1}2.04 & 0.18\\
            & S & 1.56     & 0.26 &  4.74    & 1.01 &  6.5 & 0.3   & \phantom{1}3.48 & 0.82 & \phantom{11}4.53 & 0.40 & \phantom{1}1.93 & 0.16\\
\hline \hline \\
\end{tabular}
\end{table*}

\section{Discussion}
\label{sec:discussion}

\begin{figure}[t!]
\resizebox{\hsize}{!}{\epsfig{file=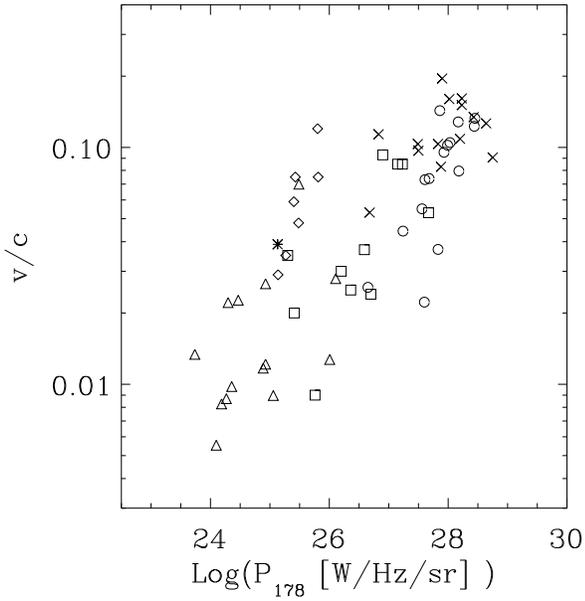}} 
\caption{\label{fig:p-v-plot} The radio power against the lobe
advance velocity for several samples of sources. Crosses: Liu et
al. (1992); Circles: Leahy et al. (1989); Squares: Alexander \& Leahy
(1987); Triangles: Parma et al. (1999); Diamonds: GRGs from this
paper; Star: The GRG B\,0313$+$683 from Schoenmakers et al. (1998).}
\end{figure}

\subsection{The relation between lobe velocity and linear size}
\label{sec:velocity-size}

In Sect. \ref{sec:velocities} we have derived the lobe advance
velocities of seven of the GRGs in our sample.  We have compared these
velocities to those found in three samples of powerful FRII-type 3CR
radio sources by Alexander \& Leahy (1987), Leahy et al.  (1989) and
Liu et al. (1992), and in a sample of lower luminosity sources
presented by Parma et al. (1999). The sizes and radio powers of the
sources in Leahy et al. and Parma et al. have been converted to a
Hubble constant of 50~km\,s$^{-1}$\,Mpc$^{-1}$.  Of the Parma et
al. sample we have only used sources which are not of FRI-type and
which show a steepening of the radio spectrum from the hotspot to the
core (`type 2' sources, in their notation). Also, the radio powers
have been converted from 1.4 GHz to 178 MHz using a spectral index of
$-0.8$.  The sample of Alexander \& Leahy consists of radio sources
with an angular size above $45\arcsec$, that of Leahy et al. of
sources larger than $20\arcsec$ and that of Liu et al. of sources
larger than $4\arcsec$. The sources 3C\,154 and 3C\,405 (Cyg A) are
studied by both Alexander \& Leahy and Leahy et al.; we have used the
better constrained values of the latter group. For all sources we have
used the average age and velocities of the two lobes.  For the GRGs we
have taken the velocities derived using the JP ageing model, since this
model has been used by Alexander \& Leahy, Leahy et al. and Liu et
al. Lastly, we have added the advance velocity of the GRG
B\,0313$+$683 (Schoenmakers et al. 1998), which has been derived in a
similar way as the GRGs presented here.

\begin{figure*}[tb]
\resizebox{0.33\hsize}{!}{\epsfig{file=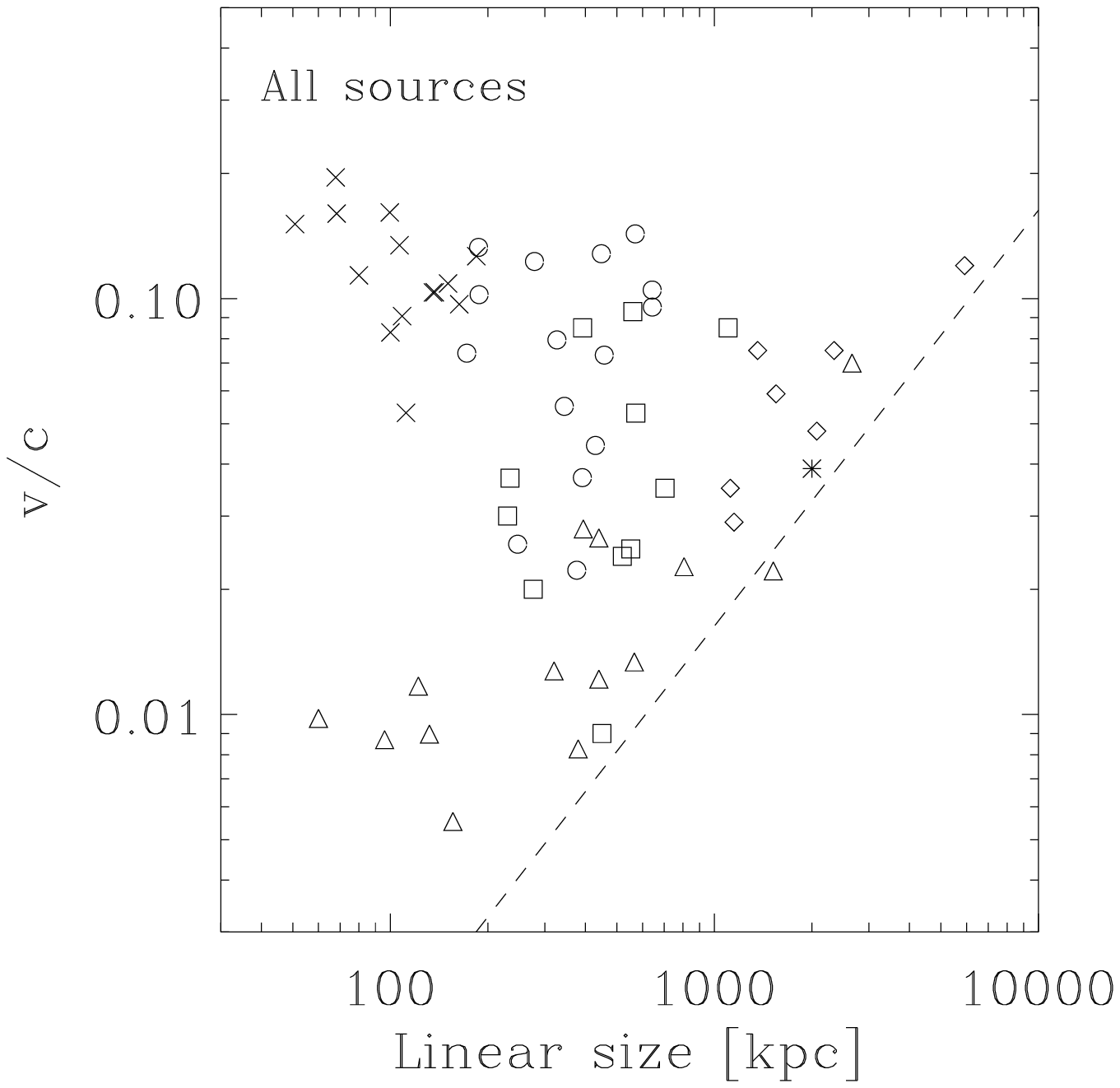}}
\resizebox{0.33\hsize}{!}{\epsfig{file=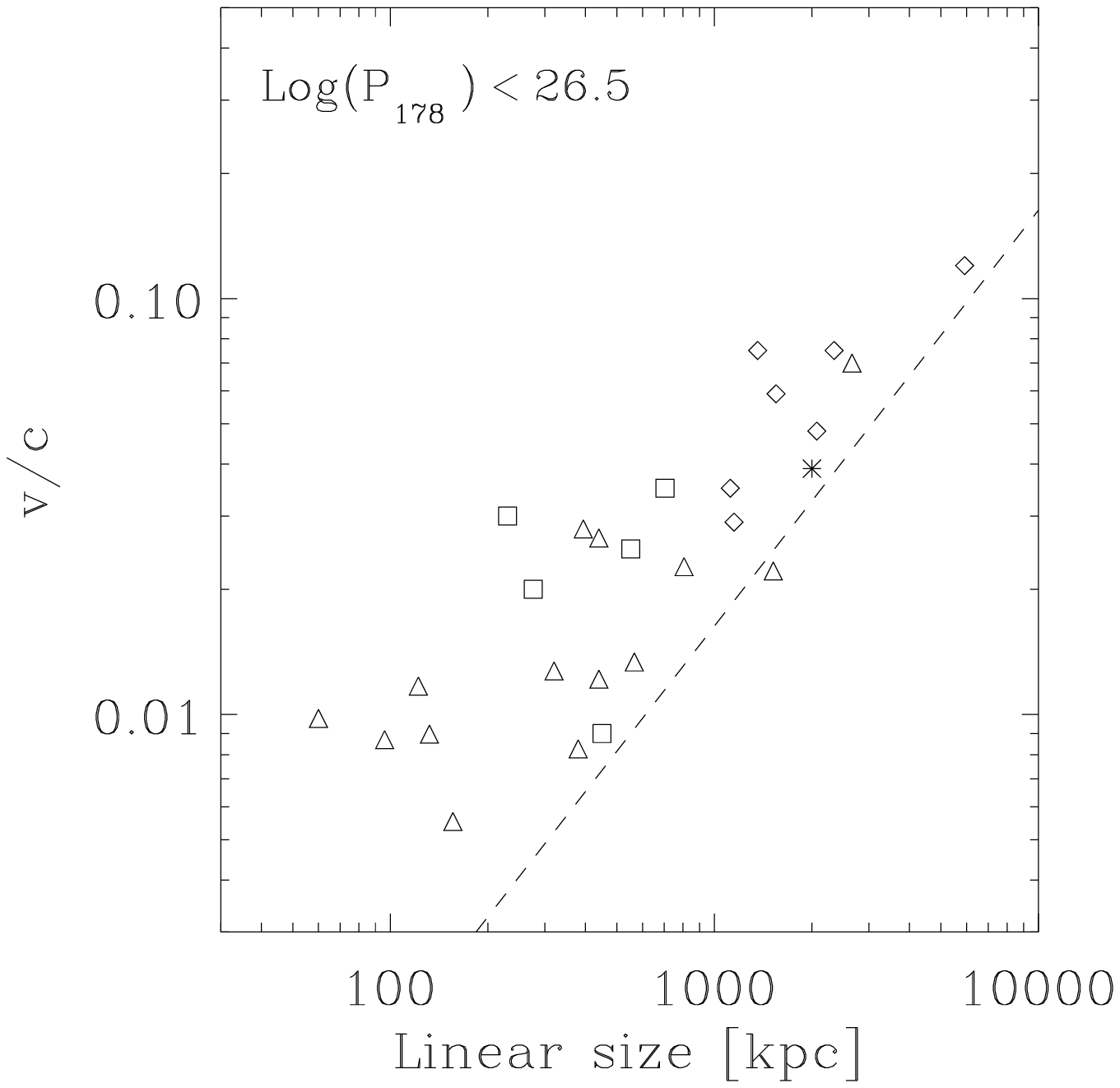}} 
\resizebox{0.33\hsize}{!}{\epsfig{file=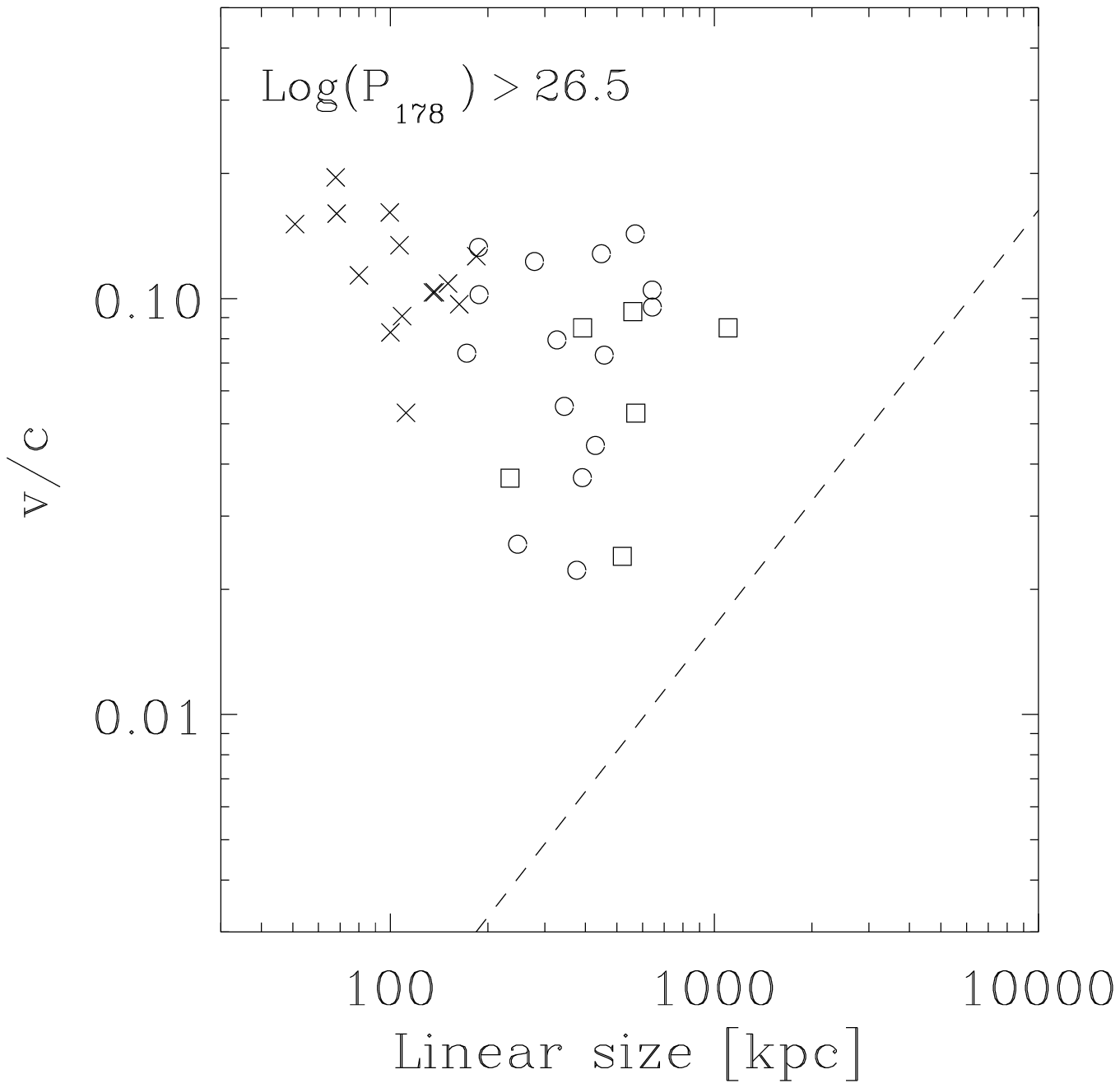}}
\caption{\label{fig:d-v-plots} The linear size of the radio sources
against the lobe advance velocities. Symbols as in
Fig. \ref{fig:p-v-plot}. The dashed line indicates the size a source
can attain at an age of $10^8$ yr. {\bf a:} (left) All sources. {\bf b:}
(middle) Sources with 178-MHz radio powers below
$10^{26.5}$\,W\,Hz$^{-1}$\,ster$^{-1}$, only. {\bf c:} (right) Sources
with 178-MHz radio powers above
$10^{26.5}$\,W\,Hz$^{-1}$\,ster$^{-1}$.}
\end{figure*}

If we plot the radio power at 178 MHz against the lobe advance
velocity for all sources (Fig. \ref{fig:p-v-plot}) we find a strong
correlation between these parameters. This correlation was noted
before by Alexander \& Leahy (1987) and confirmed by Liu et
al. (1992). We find that the GRGs agree with the correlation, but that
they have the highest advance velocities for their radio powers.  
To find out whether the advance velocity of the radio lobes is
related to the linear size of a source, we have plotted these
parameters for all sources in Fig. \ref{fig:d-v-plots}a. No
correlation is found, although the spread in velocities appears to
decrease with increasing linear size. However, when we separate the
low luminosity from the high luminosity sources, we find a strong 
correlation for the low-luminosity sources (see Fig. \ref{fig:d-v-plots}b). 
We have set the separation at a radio power of
$10^{26.5}$\,W\,Hz$^{-1}$\,ster$^{-1}$, which is in between the highest
radio power in the low-luminosity sample of Parma et al. (1999) and
the lowest in the high-luminosity samples of Liu et al. (1992) and
Leahy et al. (1989). For sources with higher radio powers
the relation between velocity and source size seems to be inverted,
although the correlation is much weaker (Fig. \ref{fig:d-v-plots}c).  

At first sight, the correlation between linear size and lobe advance
velocity for the low-luminosity sources suggests that radio sources
increase their lobe advance velocity as they increase in
size. However, the correlation can also be the result of two selection
effect, provided that radio sources decrease in radio power as they
increase in size and age (cf. Kaiser et al. 1997, Blundell et
al. 1999). First, assuming that the radio power is proportional to the jet
power and decreases with increasing linear size, large radio sources
must have higher jet powers than small radio sources of similar
observed radio power.  Therefore, that fraction of large radio sources
which are in a flux density limited sample must have the highest jet
powers of all. If all radio sources are in similar environments, a
higher jet power will most likely lead to a faster lobe advance; this
yields the observed correlation. Second, since radio sources have a
limited life-time, sources with low velocities will never become
as large as sources with high lobe velocities. This is indicated by the
dashed lines in Fig. \ref{fig:d-v-plots}, which represents an age of
$10^8$ yr; not many sources are found to the right of this line. 
Of course, sources may also live longer than this, 
but they then have to drop
below the sensitivity limit in order to disappear from the plot.   

The above cannot explain the behaviour observed for the high power
radio sources, though. There are several possible reasons for this,
among which the following (see also Parma et al. 1999).  First, the
velocity estimates in these sources depend strongly on the assumed
magnetic field strength in the lobes. This was taken as the
equipartition field strength, but as shown by Liu et al. (1992),
lowering this will drastically decrease the estimated velocities.
Second, there may be strong backflows present in the lobes of the
powerful sources (see also Scheuer 1995). This will also lead to an
overestimation of the advance velocities since one actually measures
the separation velocity between the head of the lobe and the material
flowing back. Further, the mixing of old material with much younger,
backflowing material near the radio core will lead to a further
underestimation of the age of the source, and thus to an
overestimation of the growth speed.  Scheuer (1995) suggests that backflows
are important in powerful sources such as present in the sample of Liu
et al. (1992). It is unknown if strong backflows occur in less
powerful sources as well. Lastly, in a flux density limited sample such as
the 3CR, high-power sources are at higher redshifts than low power
sources. Blundell et al. (1999) point out that this leads
to the selection of younger and smaller radio sources with increasing
redshift. Assuming that the radio power is directly related to the jet
power, and that this is directly related to the advance velocities of
the lobes, the lobes of powerful radio sources must have high
velocities. The low-power sources discussed above are all at low
redshift, so that this selection effect is not in operation.

\subsection{Spectral ages of radio sources}

\begin{figure*}[tb]
\resizebox{0.33\hsize}{!}{\epsfig{file=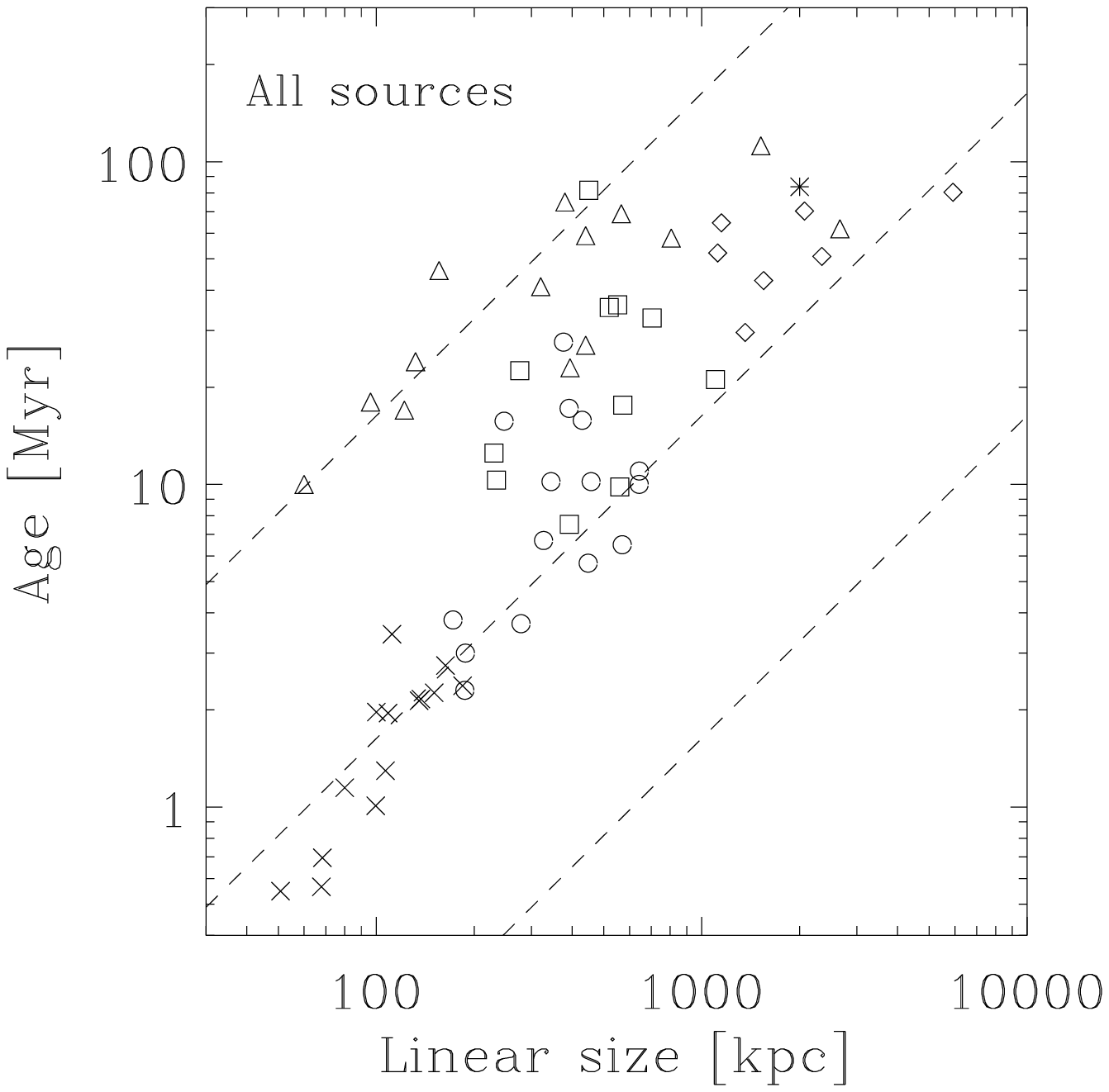}}
\resizebox{0.33\hsize}{!}{\epsfig{file=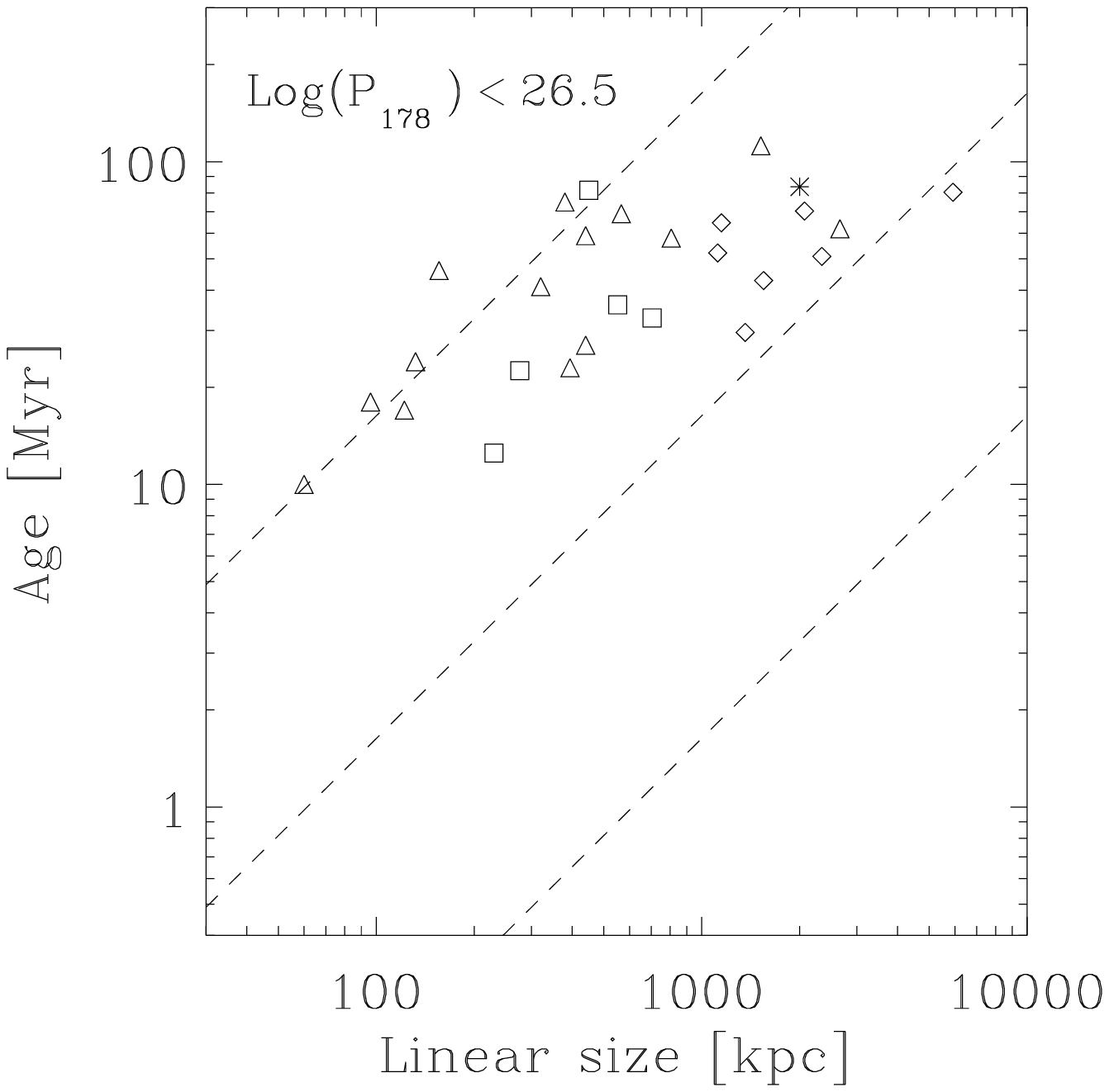}} 
\resizebox{0.33\hsize}{!}{\epsfig{file=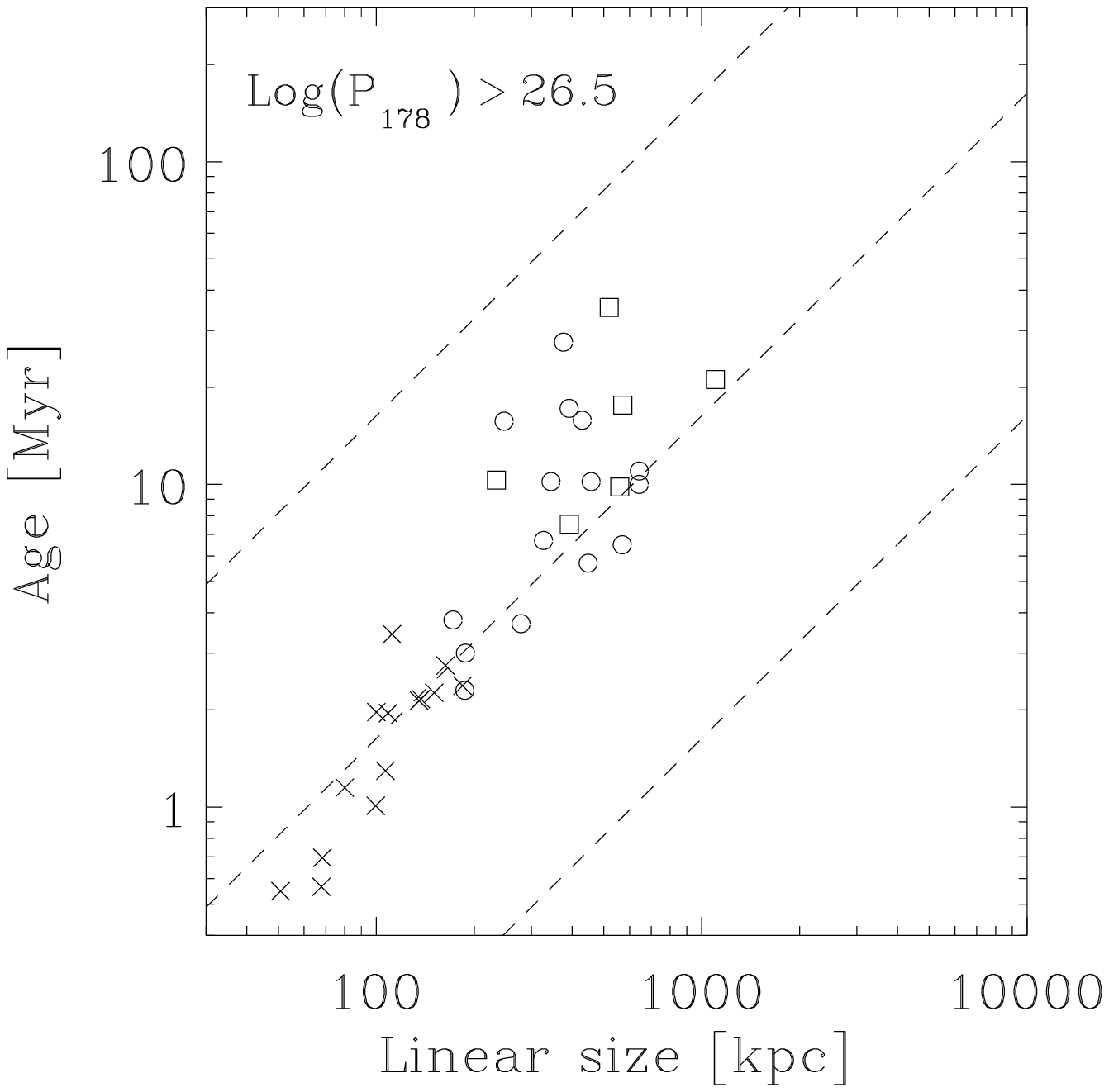}}
\caption{\label{fig:d-t-plots} The linear size of the radio sources
against their spectral age. The symbols are as in
Fig. \ref{fig:p-v-plot}. {\bf a:} (left) All sources. {\bf b:} (middle)
Sources with 178-MHz radio powers below $10^{26.5}$\,W\,Hz$^{-1}$\,ster$^{-1}$,
only. {\bf c:} (right) Sources with 178-MHz radio powers above
$10^{26.5}$\,W\,Hz$^{-1}$\,ster$^{-1}$. The diagonal lines
indicate the relation for an expansion velocity of $0.01c$ (upper
line), $0.1c$ (middle line) and $1c$ (lower line).}
\end{figure*}

Parma et al. (1999) find that the spectral ages of low power radio
sources are well correlated with their linear sizes. In
Fig. \ref{fig:d-t-plots}a we have plotted the linear sizes and
spectral ages of the sources from the samples introduced in
Sect. \ref{sec:velocity-size}. We confirm the correlation between
linear size and age, but we also find that the high-power sources have
a stronger dependence on the linear size (see
Fig. \ref{fig:d-t-plots}b and c).  This is due to the higher lobe
velocities, the cause of which we have discussed in
Sect. \ref{sec:velocity-size}.  We find that the GRGs are among the
oldest sources in this plot, although their ages are not extreme as
compared to the sources from Parma et al. This is related to the
correlation between linear size and lobe velocity (see
Fig. \ref{fig:d-v-plots}b).  Only large sources with high jet powers
will end up in our GRG sample, which probably implies that their lobes
advance faster (see previous section). This will introduce a
bias in our sample towards sources which have higher
advance velocities and thus lower ages. A multi-frequency study of
GRGs at lower flux densities (such as those presented in paper I) 
would be valuable to test the strength of this bias.

\subsection{Homogeneity of the Mpc-scale enviroment of GRGs}
\label{sec:homogeniety}

In Sect. \ref{sec:morphology} we have presented the armlength and
bending angle asymmetries of the GRGs and that of a sample of smaller-sized
3CR sources from Best et al. (1995). We find that the GRGs are
slightly more asymmetric in their armlengths, although the bending
angle distributions are similar.  The radio lobes of GRGs have
expanded well out of any extended emission-line gas regions, which
may contribute significantly to the observed distribution of
armlength-asymmetries in smaller sized radio sources (e.g. McCarthy et
al. 1991). Therefore, it can be expected that the asymmetries are
dominated by orientation effects.  If, however, the asymmetries in
GRGs are environmental, they provide information on the homogeneity,
on Mpc-scales, of the intergalactic medium (IGM) surrounding their host
galaxies.

We have attempted to fit the distribution of armlength asymmetries for
the 3CR sources and the GRGs (see Fig. \ref{fig:asymmetry}a) using
orientation only (cf. Best et al. 1995).  We find that the best fit
for our subsample of 3CR sources is obtained for a maximum angle of
$43\degr$ between the plane of the sky and the line of sight. This
assumes that the lobe velocities in a single source are equal and
distributed as determined by Best et al.  Since all sources in this
3CR subsample are radio galaxies, this agrees with the orientation
dependent unification scheme (e.g. Barthel 1989).  For the GRGs no
satisfactory fit can be found. The main reason for this is the smaller
number of sources in the first bin (see Fig. \ref{fig:asymmetry}a),
as compared to the second bin. To explain the observed distribution of
the armlength asymmetries with equal advance velocities of the two
lobes would require the introduction of a minimal allowed angle of
$\sim 14\degr$ between the radio axis and the plane of the sky.  We
therefore conclude that orientation effects alone cannot explain the
observed distribution of armlengths in GRGs.

Best et al. (1995) show how the presence of non-linearity in radio
sources alters the distribution of armlength asymmetries.  For the
GRGS, however, the incorporation of the bending angles presented in
Tab. \ref{tab:morphologies} does not improve the fit significantly.  As
shown by Best et al., bending angles have the strongest effect on the
observed armlength asymmetry distribution for sources oriented well
away from the plane of the sky. For sources close to the plane of the
sky, an increase of the maximum bending angle shifts the distribution
of armlength asymmetries more towards $x=0$, which is opposed to what
we observe.

These results suggest that the observed asymmetries must, at least
partly, be due to asymmetries in the lobe advance velocities of the
GRGs. This does not necessarily imply the existence of large-scale
inhomogeneities in the IGM. Other possibilities are a difference in
the efficiency of the energy transport down the jet, or of the jet
opening angle. Both can influence the thrust produced by the jet
material at the head of the lobe which influences the advance velocity.

In Sect. \ref{sec:morphology} we have investigated if the
armlength asymmetry is related to the lobe flux density and spectral
index asymmetries. Our most significant result is that
the shorter arm tends to have the steepest radio spectrum. Can this be
explained by environmental asymmetries?  If there is a higher density
on one side of the source, the radio lobe on that side may be slowed
down and better confined than its counterpart on the other side.
Provided that the magnetic field is frozen into the plasma, adiabatic
expansion of a radio lobe results in a decrease of the magnetic and
particle energy density (since $B\propto (r/r_0)^{-2}$ and $E \!\propto
\! (r/r_0)^{-1}$, where $r_0$ is the initial radius and $r$ the radius
after expansion, assuming spherical expansion; Scheuer \& Williams
1968).  This implies that the particles responsible for the emission
at a fixed frequency must have higher energies themselves (emitted
frequency $\propto E^2 B$), of which there are less available.  These
two mechanisms lead to a shift of the emitted spectrum to both lower
intensities and lower frequencies. Since spectral ageing itself
produces a convex spectral shape, the shift of the spectrum as a whole
to lower frequencies leads to a steeper observed spectrum in case of
an adiabatically expanded lobe as compared to a lobe which has not
expanded, or by a lesser amount (e.g. Scheuer \& Williams 1968). This
is contrary to what we find for the GRGs, arguing against a purely
environmental origin for the shortest arm to have the steepest spectrum.

A second possibility is that maybe not the expansion of the lobe as a
whole is important, but only the expansion of the radiating plasma as
it leaves the overpressured hotspot area. In a higher density
environment, and thus presumably higher pressure environment, the
pressure in the hotspot will be larger as well (strong shock
condition).  If the pressure in the lobes is roughly equal in both
lobes, then the drop in pressure of the plasma leaving the shock (or
hotspot) will be larger on the side with the higher ambient
density. Hence, the plasma will undergo a larger amount of
expansion. This expansion will be relatively fast and will thus
closely resemble the case of adiabatic expansion described above,
i.e. the somewhat aged particle energy spectrum will be shifted
towards lower energies. This will result in a steeper lobe spectrum on
the side of the source with the shortest arm, as observed.

Since we have not yet enough information on the intrinsic properties
of the jets and hotspots of the GRGs, we can not conclude that there
are large scale inhomogeneities in their Mpc-scale environment. More
detailed radio, and perhaps X-ray observations are necessary.

\subsection{Evolution of the environment of GRGs}

In an adiabatically expanding Universe filled with a hot, diffuse and
uniform intergalactic medium, the pressure, $p_{IGM}$, should increase
as a function of redshift, $z$, as $p_{IGM} \!\propto\! (1+z)^5$
(e.g. Subrahmanyan \& Saripalli 1993). In order to measure this effect
using radio lobes, it is necessary to investigate which sources have
the lowest radio lobe pressures. Radio source evolution models predict
a decrease in the lobe pressure with increasing linear size
(e.g. Kaiser \& Alexander 1997), so that it can be expected that the
largest radio sources have the lowest lobe pressures.

Cotter (1998) uses a large sample of radio sources, among which are GRGs
out to redshifts of $\sim1$, and shows that indeed the energy densities in
the radio lobes of GRGs are lower than those in smaller sized radio sources. Also, he shows that the lower limit for the lobe pressures evolves so strongly
with redshift that it does not contradict a $(1+z)^5$ relation. This prompts
the question if the equipartition pressures in the radio lobes of GRGs
indeed indicate a strong evolution in their environments.

\begin{figure*}[t!]
\setlength{\tabcolsep}{0pt}
\begin{tabular}{l l}
\resizebox{0.52\hsize}{!}{\epsfig{file=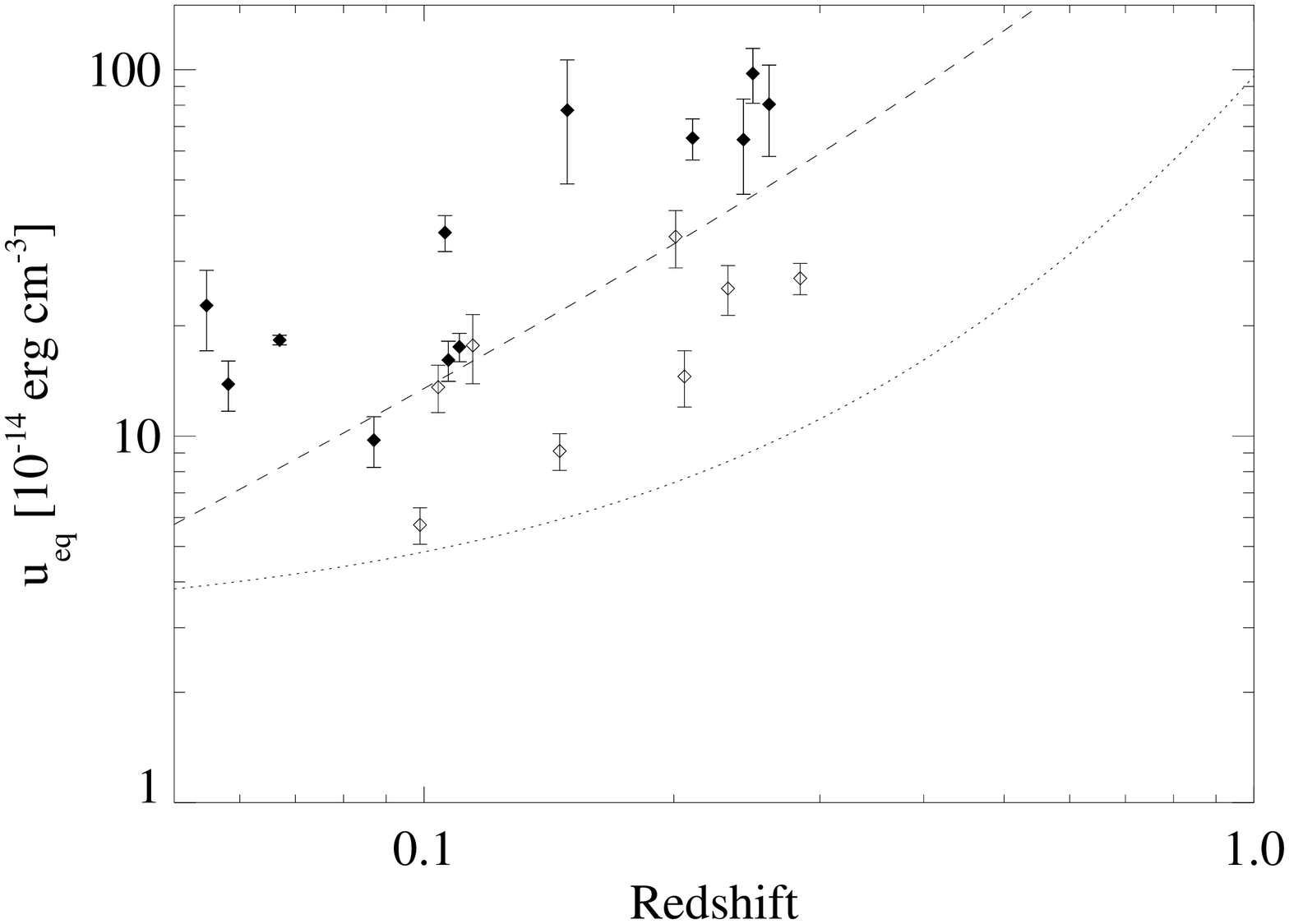,angle=90}} & 
\resizebox{0.52\hsize}{!}{\epsfig{file=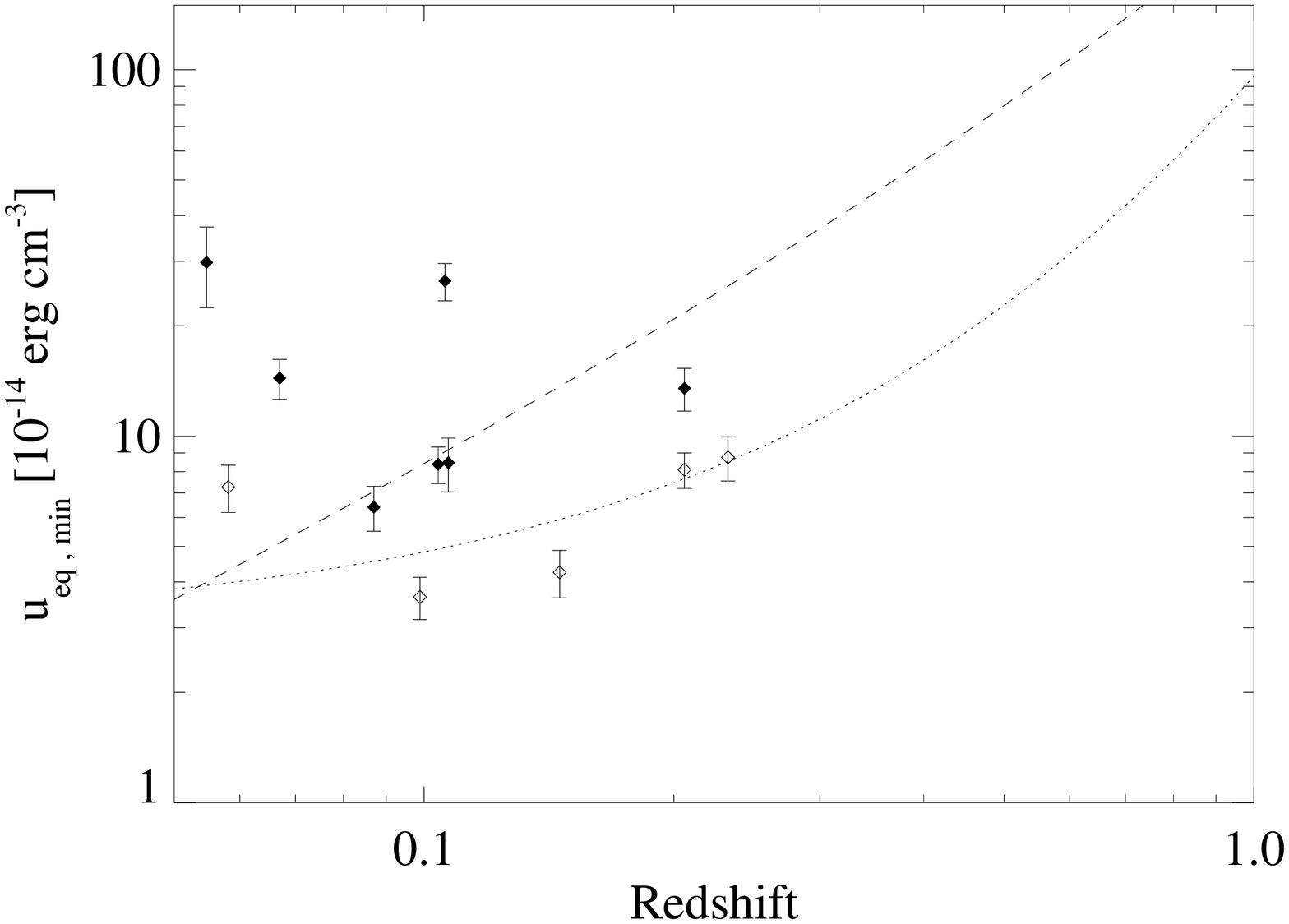,angle=90}} \\
\end{tabular}
\caption{\label{fig:ueq-z_plot} Lobe-averaged equipartition energy
densities of the FRII-type GRGs against redshift. {\bf a:} (
left) GRGs in the 1-Jy sample. Closed symbols are sources with a (projected) linear size
between 1 and 2 Mpc, open symbols are sources larger than 2 Mpc. The
dashed line indicates the expected behaviour for a source with a flux
density, volume and spectral index as given by the median values of
the sample. The dotted line indicates the lower limit if the pressure
in the lobes is dominated by relativistic particles, the lobes are in
pressure equilibrium with the IGM and the pressure, $p$, of the IGM
follows the relation $p = 1.0 \times 10^{-14}\,(1+z)^5$ dyn
cm$^{-2}$. {\bf b:} (right) The lowest measured
energy densities in the lobes (see Fig. \ref{fig:ueq-profiles})
against redshift. Symbols are as in (a).}
\end{figure*}

In Tab. \ref{tab:energy-densities} we have presented the energy
densities in the lobes of the FRII-type GRGs in the 1-Jy
sample. Although the range of redshift spanned by our sample is small
($ 0 \la z \la 0.3$), a truly strong evolution of the energy density
with redshift should show up.  In Fig. \ref{fig:ueq-z_plot}a we have
plotted for each source the average energy density of the two lobes
against its redshift. We have used different symbols for sources with
a linear size below and above 2 Mpc. The dotted line in the
figure is the expected energy density in the lobes if the IGM has a
present-day pressure, $p_{(IGM,0)}$, of $10^{-14}$ dyn\,cm$^{-2}$ as
suggested by Subrahmanyan \& Saripalli (1993), a redshift
evolution as described above and if the lobes are in pressure
equilibrium with the IGM.  We find that the energy densities in the
lobes of the sources with a size above 2 Mpc are systematically lower
than those in the lobes of the smaller sized sources, confirming the
trend, already noted by Cotter (1998), for the largest radio sources.
Also, the energy densities of the lobes of the GRGs indeed increase
strongly with increasing redshift.  We find that no source has a lobe
energy density which would place it below the line predicted by a
$(1+z)^5$ IGM pressure evolution with a present-day pressure of
$10^{-14}$ dyn\,cm$^{-2}$.

An important issue is how the selection effects affect the results.
The equipartition energy density, $u_{eq}$, is related to the total
power, $P$, and the volume, $V$, of a radio source by $u_{eq}
\!\propto\! (P/V)^{4/7}$.  The sources plotted are from a flux density
limited sample ($S \!>\! 1$ Jy at 325 MHz), which
implies that sources at larger redshift have, on average, higher radio
powers and thus higher energy densities at equal physical source
dimensions.  

In Fig. \ref{fig:ueq-z_plot}a we have plotted the expected
equipartition energy density as a function of redshift (dashed line)
for a source with a flux density equal to the median flux density of
the sample, a volume equal to the median volume and a spectral index
equal to the median spectral index of the sources.  Clearly, the slope
and position of this line follows the redshift relation of the measured
energy densities well. This suggests that the observed relation of
the energy density with redshift is most likely a result from our
use of a flux density and linear size limited sample, rather than of a
cosmological evolution.

The energy densities are the intensity weighted averages over the two
lobes. Since these values are sensitive to the contributions from the
hotspots, the values in the fainter bridges of the lobes may be
lower. From the profiles of the equipartition energy density presented
in Fig. \ref{fig:ueq-profiles} we have taken the lowest found value
for each source and plotted this in Fig. \ref{fig:ueq-z_plot}b.  We
find that most sources still lie well above the line describing the
assumed $(1+z)^5$ pressure evolution with $p_{(IGM,0)} = 10^{-14}$
dyn\,cm$^{-2}$.  Also, the dichotomy between sources smaller and
larger than 2 Mpc still exists, which indicates that even the lowest
energy densities that we measure in the lobes are systematically lower
in larger radio sources.

We can only conclude that there is no evidence in our sample for a
cosmological evolution of the energy density in the lobes of GRGs, and
that there is therefore also no evidence for a cosmological evolution
of the pressure in the IGM.  On the other hand, we are also aware that
we cannot reject the hypothesis that the pressure of the IGM evolves
as $p_{IGM} \propto (1+z)^5$, provided that the present-day
value $p_{(IGM,0)} \la 1.5 \times 10^{-14}$ dyn cm$^{-2}$.

The existence of a population of GRGs with lobe energy densities of
$\la 3 \times 10^{-14}$ erg cm$^{-3}$ at redshifts of at least 0.6
would challenge this hypothesis. Assuming a length-to-width ratio of
3, a size of 1.5 Mpc and a spectral index of $-0.8$, this requires
that GRGs have to be found with a flux density of $\la 200$ mJy at 325
MHz at these redshifts, which is well above the sensitivity limit of
WENSS and thus feasible. The flux density limit scales with the aspect
ratio, $R$, as $R^{-3}$ and will thus rapidly drop below the survey
limit of $\sim 25$ mJy (Rengelink et al. 1997) for more elongated
sources.  The only published sample of higher redshift GRGs is that of
Cotter et al. (1996), which has been selected at 151 MHz with flux
densities between 0.4 and 1 Jy. These flux densities are higher than
the above limit. The lobe pressures for the sources in this sample are
presented by Cotter (1998). Indeed, he finds an apparent evolution of
the lobe pressures in this sample, but this is therefore most likely
also the result of the use of a flux density and linear size limited
sample and not due to a true cosmological evolution of the IGM
pressure.

Lastly, we have to mention the possibility that the environments of
GRGs may be particularly underdense with respect to the environments
of other radio sources. This has sometimes been suggested as the cause
of the large size of the GRGs. If also the pressures in the ambient
medium of GRGs indeed are typically lower, then the selection of radio
sources as large as GRGs would be strongly biased towards regions with
low pressures. Since this must be valid at all redshifts, we may be
sampling the lowest pressure regions of the Universe only, which may
be untypical of the general IGM.

\section{Conclusions}
\label{sec:conclusions}

We have studied the radio properties of a complete sample of 26 giant
radio galaxies. These have been selected from the WENSS survey at a
flux density above 1 Jy at 325 MHz and an angular size above
$5\arcmin$. We have presented new radio observations of 18 of these
sources at a frequency of 10.5 GHz, obtained with the 100-m Effelsberg
telescope. These have been used in combination with available data
from the WENSS, NVSS and GB6-surveys for a low-resolution
(arcmin-scale) multi-frequency radio investigation of the properties
of these sources. We have found the following:

\begin{enumerate}
\item The armlength asymmetries of GRGs are different from those of
3CR radio galaxies. The GRGs tend to be more asymmetric, which cannot
be explained as an orientation effect only. We find that in 15 out of
the 20 FRII-type GRGs in our sample, the radio lobe which has the
steepest spectrum between 325 MHz and 10.5 GHz also has the shortest
arm. We conclude that this cannot be result of a difference in the
expansion-rate of the two radio lobes, thereby excluding asymmetries
in the environment as the major cause of this effect.
\item In profiles of the spectral index as a function of distance from
the hotspots in the lobes, we find significant steepening of the
spectrum away from the hotspot only in a few cases. Fitting these
profiles with model spectra yields a typical advance velocities of
$\sim\!0.04c$, and a spectral age of 80 Myr. Such large ages agree
with what has been found for several other GRGs in the past.
\item We find a dichotomy between powerful ($P_{178} > 10^{26.5}$
W\,Hz$^{-1}$\,ster$^{-1}$) and less powerful sources when we compare
their lobe advance velocities, as deduced from spectral ageing
studies, with their linear size.  First, we find that for linear sizes
around 100 kpc, the high power sources typically have much higher lobe
velocities than the low power sources. This dichotomy disappears for
larger sources.  Further, less powerful sources show a strong
correlation between source size and lobe advance velocity, which
extends all the way to the largest sources, the GRGs.  This may be
largely due to the following: Slowly advancing sources may never grow
out to Mpc sizes within the lifetime of the radio active phase of the
AGN.
\item Using the measured advance velocities, ages and energy contents
of the lobes of the GRGs, we find a typical particle density in front
of the lobes of a few times $10^{-5}$ cm$^{-3}$. This is in agreement
with earlier results on the density around the lobes of GRGs using
similar methods (Mack et al. 1998). Assuming a temperature of a few
times $10^6$ K, the thermal pressure in a medium with a particle
density of $4\!\times\! 10^{-5}$ would be $\sim\!2\times 10^{-14}$ dyn
cm$^{-2}$. This lies below the typical lobe pressures we find from
equipartition arguments. Profiles of the equipartition energy density
along the radio axis in the lobes indicate that there often is a
strong pressure gradient in the lobes, with the hotspots having the
highest pressures. Also this indicates that the radio lobes are
overpressured with respect to their environment.
\item The lobe pressures show a strong correlation with redshift. This
has been noted before by, e.g., Subrahmanyan \& Saripalli (1993) and
Cotter (1998). We show that the correlation in our sample can be
explained by two effects: The use of a flux density and linear size
limited sample and the method by which the equipartition lobe
pressures are calculated.  We find therefore no evidence for a
cosmological evolution of the IGM pressure between $z=0$ and
$z=0.3$. Our observations agree with a present day value of the IGM
pressure of $\sim 10^{-14}$ dyn cm$^{-2}$.
\end{enumerate}
\noindent
This has been the first study of its kind employing a complete and
relatively large sample of GRGs. The main result is that on basis of
the data presented here we find that GRGs are both old sources, in
terms of their spectral age, and that they are situated in a
relatively low density environment, but also that neither of these two
properties are extreme. Therefore, based on the study presented here, their large size probably results from a combination of these properties.


\begin{acknowledgements}
KHM was supported by the Deutsche For\-schungs\-gemein\-schaft, grant
KL533/4--2 and by the European Commission, TMR Programme, Research Network
Contract ERBF\-MRXCT96-0034 ``CERES''.  The Westerbork Synthesis Radio
Telescope (WSRT) is operated by the Netherlands Foundation for
Research in Astronomy (NFRA) with financial support of the Netherlands
Organization for Scientific Research (NWO).  The National Radio
Astronomy Observatory (NRAO) is operated by Associated Universities,
Inc., and is a facility of the National Science Foundation (NSF).
This research has made use of the NASA/IPAC Extragalactic Database
(NED) which is operated by the Jet Propulsion Laboratory, California
Institute of Technology, under contract with the National Aeronautics
and Space Administration. We acknowledge the use of NASA's {\it
SkyView} facility (http://skyview.gsfc.nasa.gov) located at NASA
Goddard Space Flight Center.\\
\end{acknowledgements}



\begin{appendix}

\section{Radio maps of the 10.5-GHz Effelsberg observations}
\label{sec:Effelsberg-maps}

Here we present the radio maps resulting from the 10.5 GHz Effelsberg
observations of 18 sources in our 1-Jy sample of Giant Radio sources
(Figs. \ref{fig:eff_1} -- \ref{fig:eff_4}).  For all sources, two
maps have been presented. The first, which is indicated by an `I' and
has the name of the source on the plot as well, shows contours of
total intensity, with $E$-field polarization vectors overplotted. 
The length of the vectors
is proportional to the polarized intensity. The scale used is
$1\arcsec\!=\!0.1$ mJy\,beam$^{-1}$, except in the case of
B\,0157$+$405 where $1\arcsec\!=\!0.05$ mJy\,beam$^{-1}$. The second
plot, indicated by `P', shows contours of polarized intensity with
$E$-field polarization vectors overplotted. Here, the length of the
polarization vectors is proportional to the fractional
polarization. The scale used is $1\arcsec = 1\%$ fractional
polarization, except in the source B\,0309$+$411, where
$1\arcsec\!=\!0.1\%$ fractional polarization. All contours are drawn at
intervals of a factor of $\sqrt{2}$, with the value of the lowest
contour given in Tab. \ref{tab:eff_contours}. In the case of
B\,0309$+$411, contours above 16 times the lowest contour are drawn at
intervals of a factor of 2. Dashed contours indicate negative flux
densities; only the first negative contour has been drawn.  References
to similar 10.5-GHz Effelsberg maps of the remaining eight sources in
the 1-Jy sample are presented in the table.

\begin{table}[t]
\setlength{\tabcolsep}{5pt}
\caption{\label{tab:eff_contours} 
Flux density values of the lowest contours in the contour plots of the
Effelsberg 10.5 GHz data (Figs. \ref{fig:eff_1} --
\ref{fig:eff_4}). Column 3 gives the flux of the lowest
contour used in the total intensity ($I$) plot in mJy beam$^{-1}$;
column 4 gives the flux density of the lowest contour in the polarized
intensity ($P$) contour plot in mJy beam$^{-1}$; column 5 gives the
reference to similar maps which have been presented elsewhere.}
\begin{tabular}{l l l r l}
\hline \hline \\
\multicolumn{1}{c}{(1)} & \multicolumn{1}{c}{(2)} & \multicolumn{1}{c}{(3)} & \multicolumn{1}{c}{(4)} & \multicolumn{1}{c}{(5)} \\
IAU name & Alt. name & \multicolumn{1}{c}{$I$} & \multicolumn{1}{c}{$P$} &  \multicolumn{1}{c}{Ref.}\\
\hline \\             
B\,0050$+$402 &           & 2.7 & 1.70 \\
B\,0055$+$300 & NGC\,315  &     &      & Mack et al. 1997 \\
B\,0109$+$492 & 3C\,35    & 4.5 & 0.95 \\
B\,0136$+$396 & 4C\,39.04 &     &      & Mack et al. 1994 \\
B\,0157$+$405 & 4C\,40.09 & 3.2 & 0.95 \\
B\,0211$+$326 &           & 2.7 & 0.80 \\
B\,0309$+$411 &           & 3.3 & 0.90 \\
B\,0648$+$733 &           & 3.0 & 1.00 \\
B\,0658$+$490 &           & 3.3 & 0.85 \\
B\,0745$+$560 & DA\,240   &     &      & Mack et al. 1997 \\
B\,0813$+$758 &           & 3.0 & 0.90 \\
B\,0945$+$734 & 4C\,73.08 &     &      & Klein et al. 1994 \\
B\,1003$+$351 & 3C\,236   &     &      & Mack et al. 1997 \\
B\,1209$+$745 & 4CT\,74.17& 3.0 & 0.90 \\
B\,1213$+$422 &           & 3.9 & 1.25 \\
B\,1309$+$412 &           & 3.9 & 1.55 \\
B\,1312$+$698 & DA\,340   & 2.7 & 0.90 \\
B\,1358$+$305 &           &     &      & Saripalli et al. 1996 \\
B\,1426$+$295 &           & 3.3 & 0.90 \\
B\,1450$+$333 &           & 1.8 & 0.90 \\
B\,1543$+$845 &           & 1.8 & 0.85 \\
B\,1626$+$518 &           & 3.9 & 0.85 \\
B\,1637$+$826 & NGC\,6251 &     &      & Mack et al. 1997 \\
B\,1918$+$516 &           & 3.9 & 1.20 \\
B\,2043$+$749 & 4C\,74.26 &     &      & Saripalli et al. 1996 \\
B\,2147$+$816 &           & 2.9 & 0.85 \\
\hline \hline \\
\end{tabular}
\end{table}

\section{Notes on individual sources}

\begin{description}
\item{\sl B\,0050+402:} A total intensity map has already been
presented by Gregorini et al. (1998); here, we have added the
polarization data. Both lobes are only weakly polarized.
\item{\sl B\,0109+402:} This source is also known as 3C\,35 and is a
fat double radio source (e.g. J\"{a}gers 1986). We have detected the
core. The complicated polarization structure in the northern lobe
results from a rapid change of the polarization angle with position,
causing beam-depolarization (see the higher resolution maps of
J\"{a}gers 1986).
\item{\sl B\,0157+405:} This source has been barely detected in our
observations. We find a spectral index of $-0.99 \pm 0.04$ between 325
MHz and 10.5 GHz, which is indeed rather steep.
\item{\sl B\,0211+326:} This source is well detected in both total
power and linearly polarized emission; the direction of the observed
$E$-field is perpendicular to the radio axis in both lobes, as is
usually found in the lobes of FRII-type sources.
\item{\sl B\,0309+411:} This is a strongly core dominated giant radio
source (de Bruyn 1989). Indeed, our radio map is dominated by the core
emission as well. The faint extension towards the north-east is an
artefact due to the beam.
\item{\sl B\,0648+733:} This is one of our largest sources. The angle
of the observed $E$-field is perpendicular to the radio axis in the
south-west lobe, and parallel in the north-east lobe. Considering the
high frequency of these observations, it is unlikely that this is
caused by Faraday rotation and this configuration therefore must be
internal.
\item{\sl B\,0658+490:} This source has a strange morphology (see
paper I). Here, only the compact central components are well
detected. The core is unpolarized at our sensitivity, the other two
components have an observed $E$-field parallel (middle component) and
perpendicular (western component) to the radio axis.
\item{\sl B\,0813+758:} This source displays a rather dominant radio
core at 10.5 GHz. Of the two lobes, only the western one is detected
in linearly polarized emission.
\item{\sl B\,1209+745:} The luminous central core/jet structure (see
van Breugel \& Willis 1981) is unresolved. Polarized emission has been
detected at the position of the jet and towards the southern lobe.
\item{\sl B\,1213+422:} The radio core of this source is quite bright
at 10.5 GHz. No significantly polarized features are detected.
\item{\sl B\,1309+412:} The total intensity map of this source has
already been presented by Gregorini et al. (1998); here, we display
the polarization data. Linear polarization has been detected in both
lobes, with observed $E$-fields parallel to the radio axis.
\item{\sl B\,1312+698:} This source is well detected in polarized
emission, with an observed $E$-field direction perpendicular to the
radio axis. The unresolved polarizaed feature in between the two lobes
may be a signature of a jet.
\item{\sl B\,1426+295:} The total intensity map shows the two radio
lobes and the core of this large source. The $E$-field direction of
the polarized emission in the two lobes is perpendicular to the radio
axis.
\item{\sl B\,1450+333:} This is one of the `double-double' radio
sources presented in Schoenmakers et al. (2000b). The resolution of
these observations is too low to investigate the properties of the
inner structure in detail. However, we find polarized emission towards
the inner structure. Polarized emission has also been detected towards
both outer radio lobes, with $E$-field directions perpendicular to the
radio axis.
\item{\sl B\,1543+845:} In total intensity this source looks like a
normal FRII-type radio source. In polarized emission, only the
northern lobe of this source is detected, with the $E$-field directed
perpendicular to the radio axis.
\item{\sl B\,1626+518:} The contour map shows the two lobes and the
relatively bright radio core of this source. Also polarized emission
has been detected in both lobes.
\item{\sl B\,1918+516:} Both lobes and a radio core are detected,
although the latter is somewhat confused with a nearby
source. Polarized emission has been detected towards both radio lobes,
with $E$-field directions almost perpendicular to the radio axis.
\item{\sl B\,2147+816:} The radio core and the two extended lobes of
this 3.7 Mpc large source are well detected. Polarized emission is
detected towards both radio lobes, with $E$-field directions almost
perpendicular to the radio axis. The southern lobe shows signs of a
change in the orientation of the polarization angle at its southern
edge.
\end{description}

\begin{figure*}
\caption{\label{fig:eff_1}Radio contourplots from the 10.5-GHz Effelsberg observations of the sources B\,0050+402, B\,0109+492, B\,0157+405, B\,0211+326, B\,0309+411 and B\,0648+733.}
\end{figure*}

\begin{figure*}
\caption{\label{fig:eff_2}Radio contourplots from the 10.5-GHz
Effelsberg observations of the sources B\,0658+492, B\,0813+758,
B\,1209+745, B\,1213+422 and B\,1309+412.}
\end{figure*}

\begin{figure*}
\caption{\label{fig:eff_3}Radio contourplots from the 10.5-GHz
Effelsberg observations of the sources B\,1312+698, B\,1426+295,
B\,1450+333 and B\,1626+518.}
\end{figure*}

\begin{figure*}
\caption{\label{fig:eff_4}Radio contourplots from the 10.5-GHz
Effelsberg observations of the sources B\,1543+845, B\,1918+516 and
B\,2147+816.}
\end{figure*}

\end{appendix}
\end{document}